\documentclass[%
 reprint,
 superscriptaddress,
 amsmath,amssymb,
 mathtools,
 aps, 
]{revtex4-2}

\usepackage{graphicx}
\usepackage{siunitx}
\usepackage[colorlinks=true,allcolors=blue]{hyperref}
\usepackage[capitalise]{cleveref}
\usepackage{layouts}
\usepackage{upgreek,textcomp}

\newcommand{\aref}[1]{Appendix~\ref{#1}}
\newcommand{\beginsupplement}{%
        \setcounter{section}{0}
        \setcounter{table}{0}
        \renewcommand{\thetable}{S\arabic{table}}%
        \setcounter{figure}{0}
        \renewcommand{\thefigure}{S\arabic{figure}}%
        \renewcommand{\theHfigure}{S\arabic{figure}}
        \setcounter{equation}{0}
        \renewcommand{\theequation}{S\arabic{equation}}%
                \renewcommand{\thesection}{S\arabic{section}}%
     }
\DeclareSIUnit{\nothing}{\relax}

\begin{document}

\title{Superconducting microsphere magnetically levitated in an anharmonic potential with integrated magnetic readout}

\author{Martí Gutierrez Latorre}
\affiliation{%
Department of Microtechnology and Nanoscience (MC2), Chalmers University of Technology, SE-412 96 G\"oteborg, Sweden\\
 }
 \author{Gerard Higgins}
\affiliation{%
Department of Microtechnology and Nanoscience (MC2), Chalmers University of Technology, SE-412 96 G\"oteborg, Sweden\\
 }
\affiliation{%
Institute for Quantum Optics and Quantum Information (IQOQI), Austrian Academy of Sciences, Vienna, Austria\\
 }
\author{Achintya Paradkar}
\affiliation{%
Department of Microtechnology and Nanoscience (MC2), Chalmers University of Technology, SE-412 96 G\"oteborg, Sweden\\
 }
 \author{Thilo Bauch}
\affiliation{%
Department of Microtechnology and Nanoscience (MC2), Chalmers University of Technology, SE-412 96 G\"oteborg, Sweden\\
 }
 \author{Witlef Wieczorek}
\email{witlef.wieczorek@chalmers.se}
\affiliation{%
Department of Microtechnology and Nanoscience (MC2), Chalmers University of Technology, SE-412 96 G\"oteborg, Sweden\\
 }%

\date{\today}

\begin{abstract}

Magnetically levitated superconducting microparticles offer a promising path to quantum experiments with picogram to microgram objects.
In this work, we levitate a $\SI{700}{\nano\gram}\sim 10^{17}\,\mathrm{amu}$ superconducting microsphere in a magnetic chip trap in which detection is integrated. We measure the particle's center-of-mass motion using a DC-SQUID magnetometer.
The trap frequencies are continuously tunable between 30 and \SI{160}{\hertz} and the particle remains stably trapped over days in a dilution refrigerator environment.
We characterize motional-amplitude-dependent frequency shifts, which arise from trap anharmonicities, namely Duffing nonlinearities and mode couplings. We explain this nonlinear behavior using finite element modelling of the chip-based trap potential.
This work constitutes a first step towards quantum experiments and ultrasensitive inertial sensors with magnetically levitated superconducting microparticles. 

\end{abstract}

\maketitle

\section{Introduction} 

Systems of levitated nano- and microparticles in vacuum \cite{ashkinOpticalLevitationHigh1976} offer extreme isolation of the particles from the environment as well as in-situ tuning of the trapping potential \cite{millenOptomechanicsLevitatedParticles2020,levitodynamics}. These platforms provide novel opportunities for realizing ultrasensitive force \cite{Gambhir_2016,Prat-Camps_2017,monteiroForceAccelerationSensing2020,Weiss2021} and acceleration sensors \cite{Goodkind1999,canavan_2002,Johnsson_2016,monteiroOpticalLevitation10ng2017,timberlake_acceleration_2019}, for studying thermodynamics in the underdamped regime \cite{Li2010,gieselerLevitatedNanoparticlesMicroscopic2018,debiossacThermodynamicsContinuousNonMarkovian2020}, for exploring many-body physics with massive objects \cite{Dholakia2010,Lechner2013,Simpson2016,Yan2022,Rieser2022}, and for exploiting rotational degrees of freedom \cite{Delord_2017,Pedriat_2021,Stickler2021,Cosimo_2022}. Recently, the center-of-mass (COM) motion of optically levitated nanospheres was cooled to the motional ground state \cite{delic_cooling_2020,magriniRealtimeOptimalQuantum2021,tebbenjohannsQuantumControlNanoparticle2021,Ranfagni2022,Piotrowski2023}, opening the possibility to perform quantum experiments with levitated nanoparticles \cite{romero-isartLargeQuantumSuperpositions2011,oriol_collapsemodels_2011,levitodynamics}. Electrically levitated nanoparticles have also seen tremendous progress towards reaching the quantum regime \cite{Fonseca_2016,Goldwater_2019,Martinetz2020,Pedriat_2021,Lorenzo2021}.

To extend quantum control from nano- to microparticles, magnetic levitation \cite{moon_superconducting_1994} has recently gained renewed interest \cite{Takahashi2017,timberlake_acceleration_2019,Vinante_2020,zheng_room_2020,Leng_2021,Brown2021,dima_carlos_2021,hoferHighQMagneticLevitation2022,marti_2022}. Magnetic levitation can be used to levitate objects of different shapes \cite{marti_2020,Navau_2021} with masses ranging from picograms to tons \cite{moon_superconducting_1994}.
It offers extreme isolation from the environment \cite{oriol,cirio_quantum_2012,Vinante_2020,zheng_room_2020,Leng_2021,hoferHighQMagneticLevitation2022} and allows for tunable potential landscapes \cite{Pino_2018}. This combination of properties makes magnetically levitated particles particularly well suited for high precision sensing of forces and accelerations \cite{Goodkind1999,canavan_2002,Johnsson_2016,jacksonkimballPrecessingFerromagneticNeedle2016,Prat-Camps_2017,timberlake_acceleration_2019}, as well as for fundamental physics experiments with picogram to microgram objects \cite{oriol,cirio_quantum_2012,Pino_2018}. Recent experimental developments in this direction include levitating micro-magnets on top of superconductors \cite{niemetz_oscillating_2000,Wang2019,timberlake_acceleration_2019,Vinante_2020,Gieseler_2020}, diamagnetic particles in strong magnetic fields \cite{Takahashi2017,Slezak_2018,zheng_room_2020,Leng_2021,Brown2021}, and superconducting microparticles in millimeter-scale superconducting magnetic traps \cite{Waarde_2016,marti_2022,hoferHighQMagneticLevitation2022}.

We pursue magnetic levitation of superconductors in a superconducting magnetic trap, since this approach promises the least intrinsic mechanical dissipation \cite{oriol,cirio_quantum_2012}. Levitated superconductors in the Meissner state do not suffer from magnetic moment drift or intrinsic eddy current damping, unlike levitated magnets  \cite{Prat-Camps_2017,timberlake_acceleration_2019,Wang2019,Vinante_2020,Gieseler_2020}.
By using a persistent current magnetic trap \cite{vanwaardeMagneticPersistentCurrent2016} the trap can be made perfectly stable, unlike systems of diamagnetic particles levitated between strong magnets \cite{Hsu2016,Slezak_2018,zheng_room_2020}. 

We levitate a superconducting microsphere in a fully chip-based system.
The chip-based approach \cite{Weinstein1995,Reichel1999,marti_2022} enables higher magnetic field gradients and trapping frequencies, as well as the potential to scale up the system to levitate multiple particles on the same chip.
We measure the particle's motion using magnetic pickup loops which are coupled to a SQUID magnetometer.
The pickup loops are integrated in the chip; this allows for precise positioning and enhanced measurement sensitivity.
In the future we will replace the SQUID by a flux-tunable superconducting microwave cavity \cite{Rodrigues2019,Zoepfl_2020,Schmidt2020,Bera2021,Luschmann2022,zoepflKerrEnhancedBackaction2023} to achieve quantum control over the COM motion of the levitated microparticle \cite{oriol,cirio_quantum_2012,Johnsson_2016,Pino_2018}.

In this work we demonstrate stable levitation of a \SI{48}{\micro\meter}-diameter (\SI{700}{\nano\gram}) superconducting microsphere over days.
We smoothly tune the particle's COM frequencies between 30 and \SI{160}{\hertz} by varying the trap current.
We observe that the COM frequencies depend on the motional amplitudes. This arises from trap anharmonicities \cite{gieselerThermalNonlinearitiesNanomechanical2013,Gieseler2014,Fonseca_2016,Setter2019,Zheng2020,Flajsmanova2020}.
The observed behavior is consistent with estimations of the trap anharmonicities extracted from finite element modelling (FEM) of our system.
In the future we will employ cryogenic vibration isolation \cite{Maisonobe2018,DeWit2019} and feedback cooling \cite{rossiMeasurementbasedQuantumControl2018,magriniRealtimeOptimalQuantum2021,tebbenjohannsQuantumControlNanoparticle2021} to reduce the motional amplitudes, then the effects of trap anharmonicities will be mitigated.

\begin{figure}[t!bhp]
    \centering
    \includegraphics{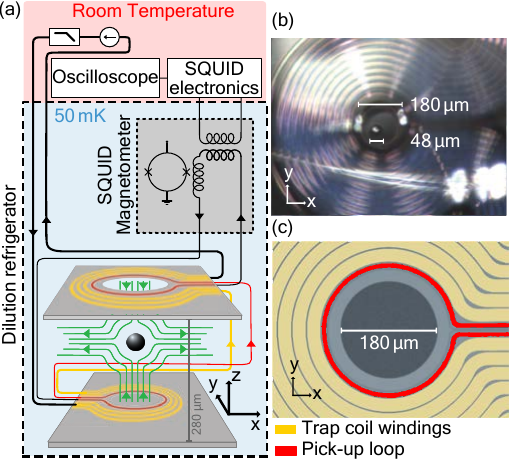}
    \caption{Chip-based magnetic levitation setup. (a)~Schematic of the experimental setup.
    The superconducting particle is levitated in a magnetic field minimum (magnetic field lines are shown in green).
    The magnetic trap coils and pickup loops are patterned on two stacked chips, which are housed in a dilution refrigerator. The vertical separation between the top and the bottom coils is \SI{280}{\micro\meter}.
    Read-out of the particle motion relies on coupling flux from the pickup loops into a SQUID magnetometer.
    (b)~Microscope image of the two-chip trap.
    Through the \SI{180}{\micro\meter} hole in the middle of the top chip, we see a \SI{48}{\micro\meter}-diameter lead microsphere resting on the bottom chip's surface.
    (c)~Scanning electron microscope image of the top view of the two-chip trap with false coloring.
}
    \label{fig:setup}
\end{figure}

\section{Experimental setup} 

The magnetic trap is produced by a current flowing in two superconducting coils, which are patterned on two silicon chips. The chips are stacked on top of each other to form an anti-Helmholtz-like configuration (see \cref{fig:setup}), for details see Ref.~\cite{marti_2022}. The superconducting particle stably levitates near the minimum of the trap's magnetic field. The particle is confined within a closed container given by the side walls of the hole in the top chip, the top surface of the bottom chip, and a glass slide on top of the hole. The force due to the magnetic trapping field is restoring within the container's bounds, with a trap depth larger than \SI{1e10}{\kelvin}. To detect the particle motion, we use two pickup loops which are integrated on the same chips, see \cref{fig:setup}.

As the particle moves it changes the magnetic flux threading the pickup loops and thus the current induced in the loops. The pickup loops are connected to a commercial DC-SQUID magnetometer, which transduces the flux into a measurable voltage. 
Typical pickup efficiencies are $\{\eta_x, \eta_y, \eta_z\}=\{1.58,3.3,19.4\}\,\mathrm{m\upphi_0}$ \si{\per\micro\meter} for the three COM modes, in terms of the flux coupled into the SQUID, and where $\phi_0$ is the magnetic flux quantum. $\eta_z >\eta_x,\eta_y$ due to the system's geometry (see \aref{app:SQUIDcalibration}).
The measurement noise floor ($0.32\,\mathrm{m\upphi_0Hz^{-0.5}}$) is limited by magnetic field fluctuations caused by trap current fluctuations and corresponds to a noise floor of  $\{200, 97, 17\}\si{\nano\meter\per\hertz\tothe{0.5}}$ for displacements along the $x, y, z$ directions, respectively.
We connect the pickup loops in series to reduce the sensitivity to these field fluctuations.
In the future, we will mitigate this noise by driving the trap using a persistent current \cite{vanwaardeMagneticPersistentCurrent2016}, which should enable reaching the intrinsic noise floor of the SQUID ($\sim 1\,$\si{\micro\nothing}$\mathrm{\upphi_0 Hz^{-0.5}}$).

The magnetic trap and the SQUID magnetometer are thermally connected to the mixing stage of a dilution refrigerator. This allows the experiment to operate between temperatures as low as \SI{50}{\milli\kelvin} and as high as the critical temperature of the superconducting particle (\SI{6.2}{\kelvin} for lead). Before levitating, the particle thermalizes on the bottom chip surface to the temperature of the chip substrate [see \cref{fig:setup}(b)]. To lift the particle off the chip surface, we ramp the trap current up to \SI{0.8}{\ampere} to overcome the adhesive force between the particle and the chip surface.
With a current of \SI{0.8}{\ampere}, the lift force is $\approx \SI{300}{\nano\newton}$, which is usually sufficient to lift the particle.
After that, the current is ramped down to the operating trap current. 

\section{Characterization of center-of-mass motion} 

Near the trap center, the particle experiences a harmonic trapping potential.
The COM frequencies depend on the trap geometry, the particle's density, and the trap current. The penetration depth ($\sim \SI{40}{\nano\meter}$ for lead) is much smaller than the particle radius (\SI{24}{\micro\meter}) and so the particle can be modelled as an ideal diamagnet with magnetic susceptibility of $-1$ and its trapping frequencies $\omega_i$ are given by \cite{Hofer_2019} 

\begin{equation}\label{eq:trapfrequency}
    \omega_i=\nabla_i B \sqrt{\frac{3}{2\mu_0\rho}}=\zeta_i \frac{\mu_0 N I}{R^2}\sqrt{\frac{3}{2\mu_0\rho}},
\end{equation}

where $\nabla_i B$ is the magnetic field gradient along the $i$ direction at the trap center, $\mu_0$ is the vacuum permeability, $I$ is the trap current, $N$ is the number of trap coil windings, $R$ is the trap coil inner radius, $\rho$ is the particle's density and $\zeta$ is a geometric factor.
At the center of an ideal anti-Helmholtz configuration $2\zeta_x=2\zeta_y=\zeta_z=0.86$.
In our system $\zeta_x =0.04 $, $\zeta_y =0.06 $ and $\zeta_z=0.12$.
(our coils are separated by \SI{280}{\micro\meter} and have inner radii $\approx \SI{125}{\micro\meter}$ as shown in \aref{app:SQUIDcalibration}). The trap axes are indicated in \cref{fig:setup}(a).

\begin{figure}[t!hbp]
    \centering
    \includegraphics[width=\columnwidth]{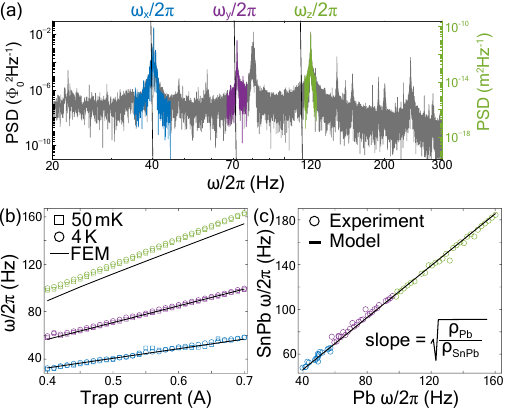}
    \caption{Trap frequencies.  (a)~Power spectrum of the SQUID signal. The frequencies of the particle's COM motion along the x, y and z directions are highlighted. The black lines show the expected trap frequencies obtained from FEM simulations. The secondary y-axis (green) shows the z amplitude in units of displacement. To obtain the x and y amplitudes in units of displacement, the secondary y-axis has to be multiplied by 151 and 35, respectively. (b)~The COM frequencies increase linearly with the trap current, as expected from \cref{eq:trapfrequency}. The results are similar when the cryostat temperature is at \SI{50}{\milli\kelvin} and at \SI{4}{\kelvin}.
    Lines show the simulated frequencies obtained from FEM.
    (c)~When the same trap settings are used, a tin-lead sphere has a higher trap frequency than a lead sphere due to its lower density (see [\cref{eq:trapfrequency}]).} 
    \label{fig:detection}
\end{figure}

\cref{fig:detection}(a) shows the power spectrum of a levitated microsphere.
Throughout this work, unless otherwise stated, we levitated a \SI{48}{\micro\meter}-diameter lead sphere using \SI{0.5}{\ampere} trap current and the cryostat temperature was \SI{4}{\kelvin}.
The peaks corresponding to the COM motion are colored.

We identified these modes by comparing the peak frequencies with predictions from FEM simulations of our system \cite{marti_2020,marti_2022}. We find good agreement between the measured and simulated COM frequencies. Peaks at the second harmonic of the COM frequencies are pronounced, particularly the second harmonic of the $\sim$\SI{40}{\hertz} peak at $\sim$\SI{80}{\hertz}.
These peaks arise from the nonlinear pickup efficiency, rather than due to actual particle motion at these frequencies. We describe these peaks further in \aref{app:harmonics}.
In the future we will feedback cool the particle, then effects of the nonlinear pickup will be negligible.

The linear relation between the COM frequencies and the trap current given by \cref{eq:trapfrequency} is shown in \cref{fig:detection}(b).
We observed no significant difference between measuring this effect with the cryostat temperature of \SI{50}{\milli\kelvin} or \SI{4}{\kelvin}.
We find good agreement between measurement results and FEM simulation results for the $x$ and $y$ modes. We attribute the $4\%$ discrepancy for the $z$-mode frequency to simulation uncertainties in the FEM.

We confirm the inverse relation between trapping frequency and particle density of \cref{eq:trapfrequency} by comparing the trapping frequencies of a lead particle and a tin-lead particle as we vary the trap current, see \cref{fig:detection}(c).
The ratio of the frequencies is given by the square root of the ratio of the material densities.

\begin{figure}[t!hbp]
    \centering
    \includegraphics[width=\columnwidth]{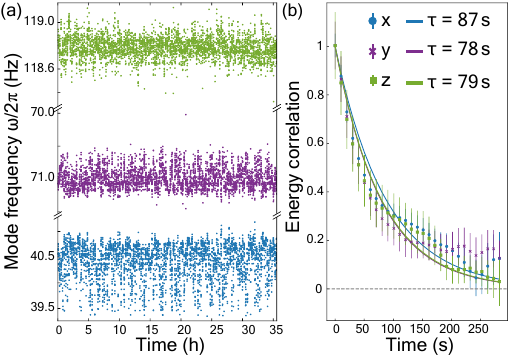}
    \caption{Trap stability. (a)~The particle stably levitates over 35\,hours and its COM mode frequencies do not drift over that time. We explain the fluctuations of the frequencies by fluctuations of the COM mode amplitudes together with frequency pulling.
    (b)~The correlations of the COM mode energies decay over around 80\,s, because of fluctuations of the mode amplitudes over this timescale. The lines represent exponential fits.}
    \label{fig:fvstime}
\end{figure}

We can stably levitate the superconducting sphere for days in the chip-based magnetic trap. \cref{fig:fvstime}(a) shows the fluctuations of the COM frequencies of a levitated sphere over a 35\,h measurement.
We have yet to observe an upper limit to the levitation time, provided that the particle is not illuminated. When the particle is illuminated, as in Ref.~\cite{marti_2022}, it heats up, loses superconductivity and falls on the bottom chip.
When the particle is kept in the dark, as in this work, we have not yet observed any upper limit to the levitation time; we have measured up to 48\,h.

Around once per day, we observe a sudden jump of all the COM frequencies of around \SI{1}{\hertz} (see \aref{app:frequencyjumps}).
We attribute these jumps to changes in the magnitude or orientation of trapped flux in the particle.
\cref{fig:fvstime}(a) shows 35\,h of data in which we do not discern any such frequency jumps.
The trapped flux will be the topic of a dedicated future investigation.

The particle's COM motion does not thermalize at the cryostat temperature, since it is strongly driven by the cryostat's vibrations. We estimate the mean amplitudes of the three modes to be $24$, $10$, and \SI{7}{\micro\meter} for the x, y, and z modes, respectively, corresponding to an effective temperature of about \SI{1e9}{\kelvin}. The energy in each mode ($m\omega_i^2\langle r_i^2\rangle$) fluctuates on a time scale $\sim 80\,\mathrm{s}$, as shown by the autocorrelation functions in \cref{fig:fvstime}(b). The mode energy fluctuations are due to the mode amplitude fluctuations, which occur over this time scale. By fitting the autocorrelation functions to exponential decay functions, we extract quality factors 3400, 4500, and 9300 for the x, y, and z modes, respectively \cite{zheng_room_2020}. We expect the quality factors to be limited by eddy current damping caused by normal-conducting metals in the vicinity of the levitated particle. In the future, we will mitigate this damping mechanism by surrounding the particle with a superconducting shield, with no normal-conducting metal within the shield.

\section{Frequency pulling} 

We can explain the COM frequency fluctuations of \cref{fig:fvstime}(a) as resulting from the fluctuating mode amplitudes [\cref{fig:fvstime}(b)] together with frequency pulling.
Frequency pulling describes the dependence of the COM frequencies on the mode amplitudes.
It arises from quartic terms in the trapping potential of the form
\begin{equation}
    U_\mathrm{pull} = \sum_i \sum_j m \gamma_{ij} r_i^2 r_j^2
\end{equation}\label{eq:frequency_pull}
which cause the motional frequencies to be shifted depending on the motional amplitudes according to \cite{bachtoldMesoscopicPhysicsNanomechanical2022}
\begin{equation}\label{eq:frequency_pull2}
    \Delta \omega_i = \frac{3 \gamma_{ii}}{\omega_i} \langle r_i^2 \rangle + \sum_{j\neq i} \frac{2 \gamma_{ij}}{\omega_i} \langle r_j^2 \rangle.
\end{equation}
The $\gamma_{ii}$ terms in the potential are called Duffing nonlinearities, while the $\gamma_{ij}$ terms describe couplings between the modes.

\begin{figure}[t!hbp]
    \centering
    \includegraphics[width=\columnwidth]{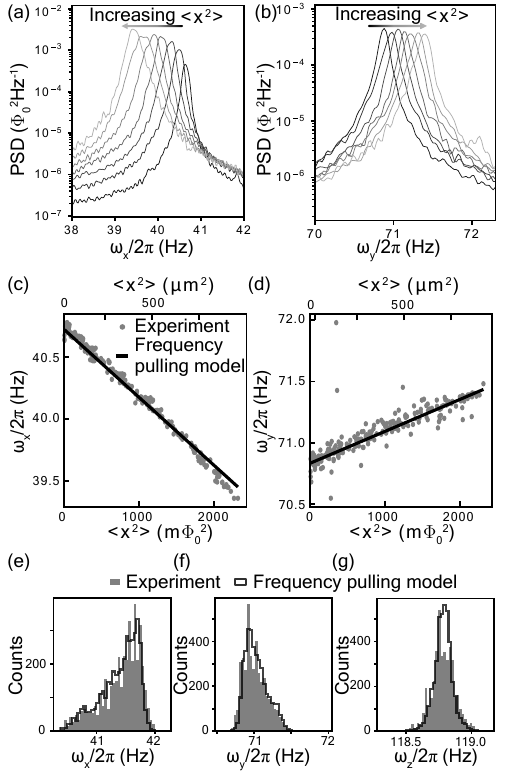}
    \caption{Frequency pulling. The frequencies of the (a) x-mode and of the (b) y-mode depend on the amplitude of the x-mode; spectra represented in lighter colors have higher x-mode amplitudes. (c) and (d)~The x-and-y mode frequencies change linearly with the mean-square amplitude of the x-mode. The slope of each model line is given by estimation of the trap anharmonicity obtained from FEM and by the estimated pickup efficiency $\eta_x$.
    (e)-(g)~The fluctuating mode frequencies of \cref{fig:fvstime}(a) are described by asymmetric distributions.
    These distributions are reproduced well by using the same frequency pulling model as in (c) and (d), together with the distribution of mode amplitudes from the experimental data.
    }
    \label{fig:amplitude}
\end{figure}

Experimental data showing frequency pulling is shown in \cref{fig:amplitude}. The spectral peak corresponding to the x-mode (y-mode) shifts depending on the amplitude of the x-mode motion in \cref{fig:amplitude}(a) [(b)].
This is described by \cref{eq:frequency_pull2}.
The remaining seven graphs showing the dependence of $\omega_i$ on $\langle r_j^2\rangle$ are included in \aref{app:pulling}.

The linear relation between the mode frequencies and the mean-square displacements [\cref{eq:frequency_pull2}] are shown in \cref{fig:amplitude}(c) and (d), in which the spectral peak frequencies $\omega_x$ and $\omega_y$ are plotted against the spectral peak area corresponding to the x-mode (the remaining graphs of the same form are included in \aref{app:pulling}).
The slopes of these lines depend on the values of $\gamma_{xx}$ and $\gamma_{yx}$ as well as the pickup efficiency $\eta_x$.
We extract estimates of $\gamma_{ij}$ from FEM simulations of our system (see \aref{app:anharmonicity}); this allows us to use the nine gradients (i.e.\ the nine linear relations between $\omega_i$ and $\langle x_j^2 \rangle$, for $i\in\{x,y,z\}$ and $j\in\{x,y,z\}$) to estimate the three efficiencies $\eta_i$ (the values are quoted earlier).
The estimated efficiencies are used to convert the lower x-axes of \cref{fig:amplitude}(c)-(d) into the upper x-axes, and yield the secondary y-axis of \cref{fig:detection}(a).

The estimation of the three efficiencies is an over-constrained problem, it yields fairly consistent results for the nine graphs. For instance, the estimation of $\eta_x$ together with the FEM estimations of $\gamma_{xx}$ and $\gamma_{yx}$ describes well both the slope in \cref{fig:amplitude}(c) and \cref{fig:amplitude}(d).

To observe the frequency pulling effect presented in \cref{fig:amplitude} we did not control the particle's motional amplitudes, instead, we filtered a long 35\,h dataset in which the motional amplitudes randomly fluctuated:
We separated the data into 10\,s chunks and extracted the mode frequencies and mode areas from each chunk.

To investigate the dependence of the mode frequencies on, e.g., the x-mode amplitude, we filtered out the data in which the y- and z-mode amplitudes were high.
Each point in \cref{fig:amplitude}(c)-(d) corresponds to one 10\,s chunk.
To produce the seven spectra in \cref{fig:amplitude}(a) and (b) we binned the power spectrum of each 10\,s-long dataset into seven bins, based on the x-peak areas, then we averaged the power spectra for each of the seven bins.

The mode amplitude fluctuations together with frequency pulling describe the frequency fluctuations of \cref{fig:fvstime}(a).
We represent the frequency distributions by histograms in \cref{fig:amplitude}(e)-(g); we note they are asymmetric.
We also plot histograms based on the frequency pulling model; the model histograms were constructed by predicting the mode frequencies for each 10\,s interval of data based on the measured mode amplitudes, the nonlinear coefficients $\gamma_{ij}$ and the pickup efficiencies $\eta_i$.
The model histograms describe the observed frequency fluctuations well.

Because the modes are coupled via the trap anharmonicity [\cref{eq:frequency_pull2}] we expect the average energies of the modes to be similar.
We can estimate the mean energies of the modes $E_i \approx m \omega_i^2 \langle r_i^2 \rangle$ using the estimated pickup efficiencies $\eta_i$, and we find $E_x = 2.1 E_y = 1.6 E_z$.
Because the estimated average energies rely on the estimated pickup efficiencies, the similarity of the average energies lends credence to our estimates of $\eta_i$.

\section{Conclusions} 

In conclusion, we have demonstrated an integrated superconducting chip device for magnetic levitation and detection of superconducting microparticles at mK temperatures. We have shown a good understanding of the trap potential, the COM motion, and inductive coupling by comparing trap models with measurements. Further, we have demonstrated that the COM frequencies can be tuned via the trap current and that the system is stable over days. For the large motional amplitudes in our experiments, we observe nonlinear behavior and mode coupling of the COM particle motion, in the form of amplitude-dependent frequency shifts. 

Future experiments will employ a cryogenic passive vibration isolation system \cite{Maisonobe2018,DeWit2019}, which will decouple the particle motion from external mechanical vibrations. Such vibration isolation systems can attenuate our measured mechanical vibrations of about \SI{1e-8}{\meter\per\hertz\tothe{0.5}} by more than six orders of magnitude to achieve thermally driven COM motion at \SI{50}{\milli\kelvin}. Further, the detection noise floor can be greatly reduced by the use of a persistent current trap \cite{vanwaardeMagneticPersistentCurrent2016} and improved magnetic shielding, which should allow us to reach the intrinsic noise floor of  $1\,$\si{\micro\nothing}$\mathrm{\upphi_0 Hz^{-0.5}}$ of our commercial SQUID. Alternatively, the particle motion can also be measured using flux-sensitive superconducting microwave resonators \cite{Rodrigues2019,Zoepfl_2020,Schmidt2020,Luschmann2022,zoepflKerrEnhancedBackaction2023}. Furthermore, the pick-up efficiency can be increased by placing the pick-up coil closer to the particle, by using multi-winding coils, and by improving the flux-transfer efficiency to the SQUID. We estimate that these measures will yield an improvement in the pick-up efficiency of four orders of magnitude. This improvement is enabled by our understanding of the chip-based particle motion detection and the flexibility in the microfabrication of pick-up loops of our chip-based approach. Overall, the reduction of technical noise and increase in pick-up efficiency should enable feedback cooling to the quantum ground state \cite{rossiMeasurementbasedQuantumControl2018,magriniRealtimeOptimalQuantum2021,tebbenjohannsQuantumControlNanoparticle2021} of the COM motion of microparticles, which would serve as a gateway to generating quantum states of nano- to microgram masses \cite{oriol,cirio_quantum_2012,Johnsson_2016}.

The data of this study are openly available in zenodo.org at \href{https://doi.org/10.5281/zenodo.7304535}{https://doi.org/10.5281/zenodo.7304535}.

\begin{acknowledgments}
We gratefully acknowledge Axel Eriksson, Andreas Isacsson, Philip Schmidt, Joachim Hofer, and Fabian Resare for insightful discussions and Joachim Hofer also for sharing of code. This work was supported in part by the Horizon Europe 2021-2027 Framework Programme under the Grant Agreement number 101080143 (SuperMeQ), the QuantERA project C’MON-QSENS!, the Knut and Alice Wallenberg Foundation through a Wallenberg Academy Fellowship (W.W.), by the Wallenberg Center for Quantum Technology (WACQT, A.P.), by Chalmers Excellence Initiative Nano, and by the Swedish Research Council (Grant 2020-00381, G.H.). Sample fabrication was performed in the Myfab Nanofabrication Laboratory at Chalmers. Simulations were performed on resources provided by the Swedish National Infrastructure for Computing (SNIC) at Tetralith, Link\"oping University, partially funded by the Swedish Research Council (Grant 2018-05973).
\end{acknowledgments}


\clearpage

\addcontentsline{toc}{section}{Supplementary Material}
\section*{Supplemental Material}

\beginsupplement


\section{Extracting the trap anharmonicity from FEM}\label{app:anharmonicity}

We consider anharmonicities in the trapping potential up to quartic terms.
The potential can then be described as
\begin{align} \label{eq_potential}
\begin{split}
    U &= \sum_i \tfrac{1}{2} m \omega_i^2 r_i^2 \\
    &+ \sum_i \sum_j \sum_k m \beta_{ijk} r_i r_j r_k \\
    &+ \sum_i \sum_j \sum_k \sum_l m \gamma'_{ijkl} r_i r_j r_k r_l
\end{split}
\end{align}
The force acting on the particle is
\begin{equation}\label{eq_force}
    F_i = -\frac{dU}{dr_i}
\end{equation} 
The FEM simulations give us the force acting on the particle due to the trapping potential at different displacements, as described in Ref.~\cite{marti_2020}.
We fit the FEM results by \cref{eq_force} with the coefficients of the trapping potential of \cref{eq_potential} as free parameters.
In this way we extract the trapping potential coefficients from the FEM simulations.

Note, in comparing $\gamma'$ with $\gamma$ from the main text:
\begin{align}
\begin{split}
    \gamma_{ii}&=\gamma'_{iiii} \\
    \gamma_{ij}=\gamma_{ji}&=\tfrac{1}{3}\gamma'_{iijj}=\tfrac{1}{3}\gamma'_{ijji}=\tfrac{1}{3}\gamma'_{jjii}\\
    &=\tfrac{1}{3}\gamma'_{ijij}=\tfrac{1}{3}\gamma'_{jiij}=\tfrac{1}{3}\gamma'_{jiji}
\end{split}
\end{align}

\begin{figure*}[t!hbp]
    \centering
    \includegraphics{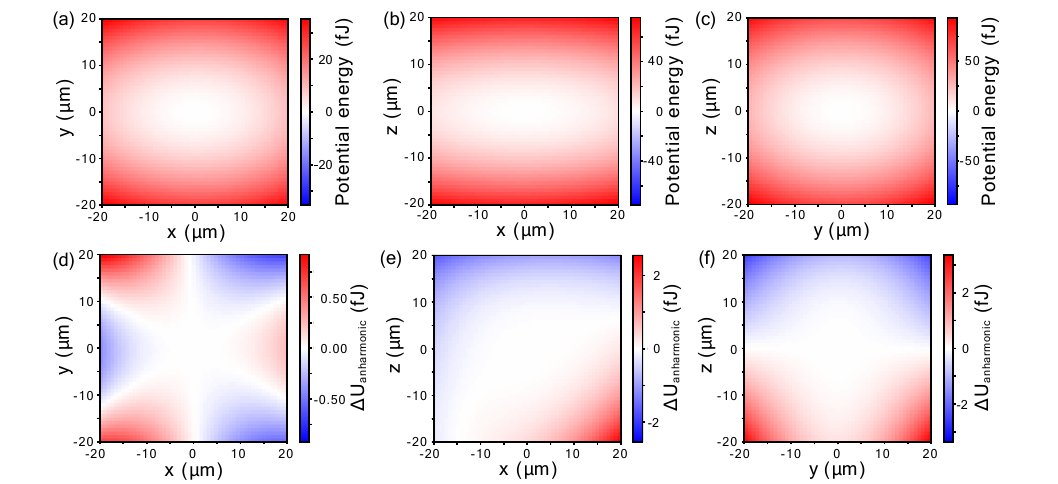}
    \caption{Model of the potential energy landscape of a \SI{48}{\mu\meter} diameter particle in the chip trap. The model coefficients are extracted from FEM. (a-c)~2D slices of the potential energy. (d-f)~Contributions from the cubic and quartic terms to the potential landscape.}
    \label{fig:anhormonic_potential}
\end{figure*}

The potential energy which we calculate based on fitting the FEM simulations is represented in \cref{fig:anhormonic_potential}. We calculate the potential energy in 3D; in the figure 2D slices are shown.
\cref{fig:anhormonic_potential} also shows the relative contribution of the cubic and quartic terms to the potential.

\begin{figure*}[t!hbp]
    \centering
    \includegraphics{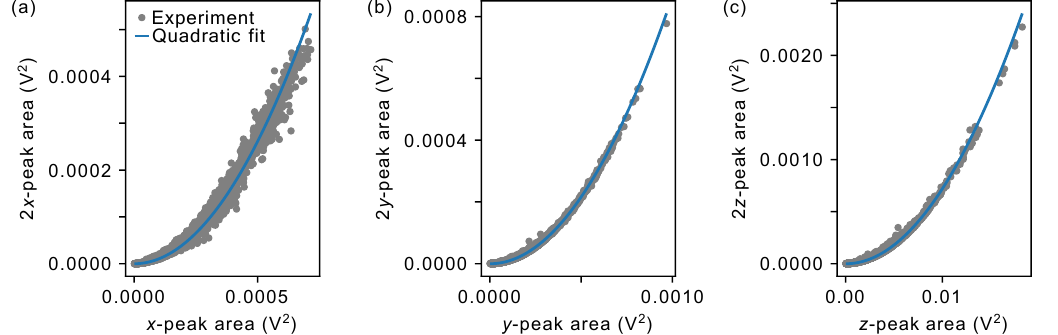}
    \caption{Comparison of fundamental and second harmonic mode spectral areas. The areas of spectral peaks at the second harmonic frequencies increase quadratically with the areas of the spectral peaks at the fundamental frequencies.}
    \label{fig:harmonic_area}
\end{figure*}

\section{Measurement of particle motion}\label{app:SQUIDcalibration}

As the particle displacement oscillates the magnetic flux threading the pickup loops oscillates in response.
This oscillating flux of the pickup loops is transferred to the SQUID loop.
The flux at the SQUID loop is converted into a voltage signal.
In this section each of these steps are described in turn.

\subsection{Inductive coupling to particle motion}\label{app:inductivecoupling}

\begin{figure}[t!hbp]
    \centering
    \includegraphics{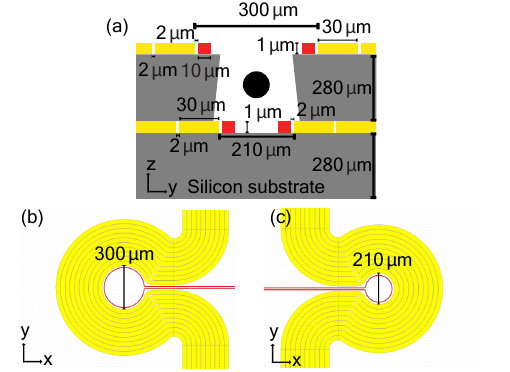}
    \caption{Chip trap. (a)~Schematic of the cross-section of the stacked two-chip trap. Layout of the top (b) and bottom (c) coils. The inner radius of the bottom coil is reduced to achieve a stronger lift force on the particle.
    }
    \label{fig:twochiptrap}
\end{figure}

\begin{figure*}[t!hbp]
    \centering
    \includegraphics{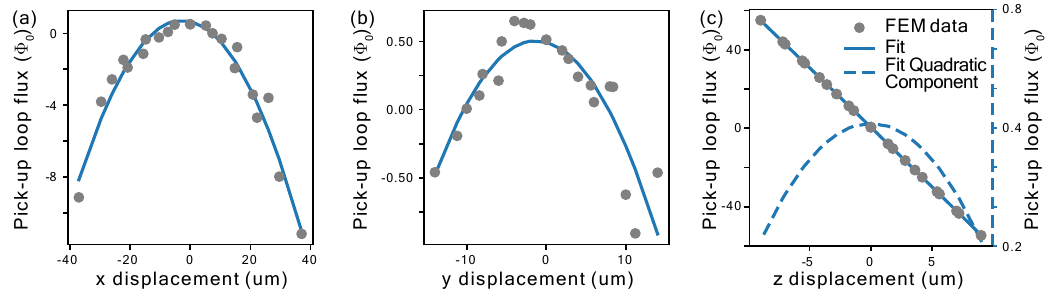}
    \caption{Dependence of flux in the pickup loop on particle displacement. The pickup loop flux responds quadratically to particle displacements along all directions. There are also linear components to the response. The response to displacement along the z direction is near-linear. The nonlinear response of the pickup loop flux leads to additional peaks in the spectrum of \cref{fig:higherorderfrequencies} and explains the behavior of \cref{fig:harmonic_area}.}
    \label{fig:pickupsignal}
\end{figure*}

We estimate the response of the change of flux threading the pickup loops in response to displacement of the particle along x, y and z by treating the trapping field as a magnetic quadrupole field (which is a good approximation near the center), using Eq.~(5) from \cite{Hofer_2019}, and using knowledge of the geometry of our system (represented in \cref{fig:twochiptrap}).
The results are shown in \cref{fig:pickupsignal}.
The response to displacement along all directions is nonlinear. That is to say, the response has strong quadratic components ($\{-7.5, -6.0, -3.7\}\,\mathrm{m\upphi_0}$\SI[parse-numbers = false]{}{\per\micro\meter\squared}) which we describe by the parameter $u_i$ in \cref{app:pickupnonlinerity}. We note that the response of the pickup to displacement along the z direction is largely linear, whereas the response to x and y displacements has a clear quadratic component. The response can be made more linear along the x and y directions by changing the geometry such that the particle equilibrium position is displaced further from the center of the pickup loops.
Because the pickup loops are connected in series, the overall pickup-loop flux is the sum of the fluxes in each loop.

\subsection{Flux transfer from the pickup loops to the SQUID loop}

\begin{figure}[t!hbp]
    \centering
    \includegraphics{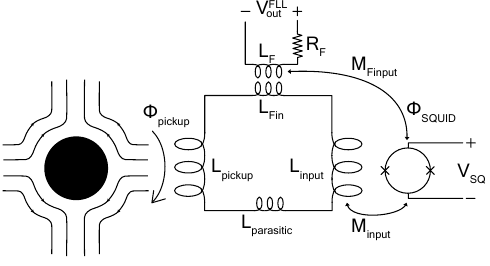}
    \caption{Schematic circuit diagram of the single stage DC-SQUID magnetometer used to detect the particle motion \cite{Drung2007}.}
    \label{fig:SQUIDelectronics}
\end{figure}

\cref{fig:SQUIDelectronics} shows how the pickup loops are electrically connected to the SQUID via an input coil.
The flux transfer is given by
\begin{equation}\label{eq:fluxtransfer}
    \phi_\mathrm{SQUID} = \frac{M_{\mathrm{input}}}{\left(L_\mathrm{pickup}+L_\mathrm{input}+L_\mathrm{parasitic}\right)} \phi_\mathrm{pickup} = \eta_{\mathrm{flux}}\cdot \phi_\mathrm{pickup} 
\end{equation}
where $L_\mathrm{input}\approx 24\,\mathrm{nH}$ is the input coil inductance, $L_\mathrm{parasitic}\approx 33\,\mathrm{nH}$ is the parasitic inductance which is dominated by the inductance of the twisted wire pairs, $L_\mathrm{pickup}\approx 0.72\,\mathrm{nH}$ is the pickup inductance, and $M_\mathrm{input}\approx 0.87\,\mathrm{nH}$ is the mutual inductance between the SQUID loop and the input coil inductance. Using these values, we estimate the flux transfer efficiency $\eta_{\mathrm{flux}} = 3.1\times10^{-2}$. By improving the inductance matching, $\eta_{\mathrm{flux}}$ can be improved to 0.5.

Based on the dependence of the pickup-loop flux on the particle displacement (\cref{fig:pickupsignal}) and the flux transfer efficiency from the pickup loops to the SQUID loop \cref{eq:fluxtransfer} we make independent estimates of the pickup efficiency in the pick-up loop $\eta_{i,\mathrm{quad-field}}^{\mathrm{pickup}}=\{40,15.3,600\}\,\mathrm{m\upphi_0}$\SI[parse-numbers = false]{}{\per\micro\meter}. These values are close to the ones found by using the measurement data and the model for the potential energy, where we obtain $\eta_{i,\mathrm{model}}^{\mathrm{pickup}}=\{50,106,625\}\,\mathrm{m\upphi_0}$\SI[parse-numbers = false]{}{\per\micro\meter}. We note that we obtain the latter values by dividing the pickup efficiencies in terms of flux in the SQUID $\eta_i=\{1.58,3.3,19.4\}\,\mathrm{m\upphi_0}$\SI[parse-numbers = false]{}{\per\micro\meter} by the flux transfer efficiency $\eta_{\mathrm{flux}}$, i.e., $\eta_{i,\mathrm{model}}^{\mathrm{pickup}}=\eta_i/\eta_{\mathrm{flux}}$. The small discrepancy may arise because our estimates of the response of the pickup loops to displacements along $x$ and $y$ using the data in \cref{fig:pickupsignal} depend critically on our knowledge of the particle's equilibrium position, due to the nonlinearity of the pickup efficiency. Additionally, we made the simplifying assumption in \cref{app:inductivecoupling} of a magnetic quadrupole field at the trap center.

\subsection{Flux in SQUID loop converted to SQUID voltage}
We operate the SQUID in flux-locked loop mode. Then the SQUID output voltage $V_\mathrm{out}$ is related to the SQUID loop flux $\phi_\mathrm{SQUID}$ by
\begin{equation}\label{eq:SQUIDflux_to_voltage}
    \phi_\mathrm{SQUID}=\frac{M_\mathrm{Finput}}{R_\mathrm{F}}V_\mathrm{out}^\mathrm{FLL}
\end{equation}
where $M_\mathrm{Finput}=$\SI{38}{\pico\henry} is the mutual inductance between the feedback electronics and the SQUID, and $R_\mathrm{F}=$\SI{10}{\kilo\ohm} describes the resistor used in the feedback electronics.

\section{Harmonics in spectra}\label{app:harmonics}

\begin{figure}[h!tbp]
    \centering
    \includegraphics{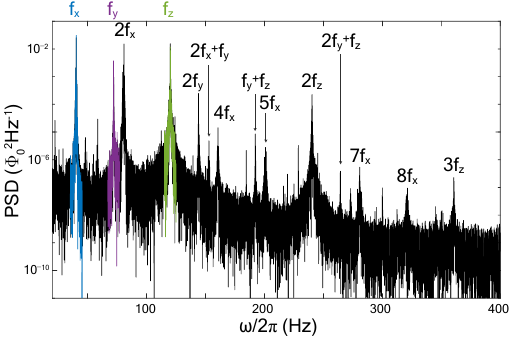}
    \caption{The power spectrum displays peaks at the harmonics and at mixing frequencies, as well as peaks at the COM motional frequencies of the particle (the fundamental peaks are colored). The harmonics arise primarily from the nonlinearity of the pickup.}
    \label{fig:higherorderfrequencies}
\end{figure}

As well as observing the peaks in the spectra at frequencies $\omega_x$, $\omega_y$ and $\omega_z$ corresponding to COM motion along the x, y and z directions, we observe strong peaks at frequencies $2\omega_x$, $2\omega_y$ and $2\omega_z$ (see the power spectrum in \cref{fig:higherorderfrequencies}).

We observe that the peak areas of the second-harmonic peaks grow quadratically with the peak areas of the respective fundamental peaks, as shown in \cref{fig:harmonic_area}.
The fundamental and harmonic peak areas $A_{\mathrm{f}i}$ and $A_{\mathrm{h}i}$ are related by
\begin{equation}
    A_{\mathrm{h}i} = R_i^{\mathrm{obs}} A_{\mathrm{f}i}^2
\end{equation}
The factor $R_i^\mathrm{obs}$ describes the observed quadratic relation between the peak areas, it is different for each mode.
From the data in \cref{fig:harmonic_area} we extract $R^\mathrm{obs}=\{1000, 870, 7.2\}\,\mathrm{V^{-2}}$ for x, y and z modes, respectively (in this section we work with the spectral peak areas in units of $\mathrm{V^2}$, that is why $R$ has units $\mathrm{V^{-2}}$).

Constructing \cref{fig:harmonic_area} involved breaking the 35\,h-long dataset which was used in Fig.~3 and Fig.~4 of the main text into 10\,s chunks, we then calculated the power spectrum for each 10\,s chunk and extracted the peak areas for the three fundamental and three second harmonic peaks and compared them.

Below, we estimate that the actual motion of the particle at frequencies $2\omega_x$, $2\omega_y$ and $2\omega_z$ arising from the trap anharmonicity is negligible.
Instead, we believe the second harmonics arise primarily due to the pickup nonlinearity.

\subsection{Particle motion at the second harmonic frequency arising from the trap anharmonicity}
In a potential of the form
\begin{equation}
    U = \tfrac{1}{2} m \omega_x^2 x^2 + \tfrac{1}{3} m \beta_{xxx} x^3
\end{equation}
the particle will move with a component at frequency $\omega_x$ and a component at frequency $2\omega_x$.
The amplitudes of these motions, at the fundamental frequency and at the harmonic frequency, are related by
\begin{equation}
    \langle x_h^2 \rangle = \frac{\beta_{xxx}^2 \langle x_f^2 \rangle^2}{18 \omega_x^4}
\end{equation}
provided that the cubic term in the potential is a perturbation and provided that the motional amplitudes are small.

We derive this equation by solving the equation of motion in this potential looking for a solution of the form
\begin{equation}
    x(t) = X_1 \sin{\omega_x t} + X_2 \sin{(2\omega_x t + \theta)}
\end{equation}
in the regime when $X_2 \ll X_1$.

Because the peak areas are related to the mean-square amplitudes by $\langle x_i^2 \rangle = \eta_i A_i$ we expect the nonlinearity of the trapping potential to give
\begin{equation}
    R_i^\mathrm{trap} = \frac{A_{\mathrm{h}i}}{A_{\mathrm{f}i}^2} = \frac{\beta_{iii}^2 \eta_i}{18 \omega_i^4}
\end{equation}

And thus, using the values $\beta_{iii}$ extracted from FEM (see \cref{app:anharmonicity}) we expect $R^\mathrm{trap}=\{6\times10^{-1},1.9\times10^{-4},2.1\times10^{-4}\}\,\mathrm{V^{-2}}$ for the x, y and z modes.
These values are much smaller than the observed values $R^\mathrm{obs}$, and so the second harmonic peaks do not arise from motion of the particle at the second harmonic frequencies.

\subsection{Second harmonic peaks caused by the pickup nonlinearity}\label{app:pickupnonlinerity}

A particle in a 1D harmonic potential moves according to
\begin{equation}
x(t)=X \sin{\omega_x t}
\end{equation}

As described in \cref{app:inductivecoupling} and \cref{fig:pickupsignal}, the flux threading the pickup loops depends nonlinearly on the particle position $x(t)$.
This can be captured by the quadratic function
\begin{align}
\phi(t) &= u x(t)^2 + v x(t) + w \\
&= \frac{u X^2}{2} (1 - \cos{2 \omega t}) + v X \sin{\omega t} + w
\end{align}
Here $v$ and $u$ are conversion factors from particle displacement to pickup-loop flux; $v$ describes the linear response of the pickup-loop, while $u$ describes the quadratic response. $w$ is an offset.

The SQUID voltage signal is related to the flux threading the pickup loops by
\begin{equation}
V(t) = k \phi(t)
\end{equation}
where we define the conversion factor $k$.
And so, the voltage signal is related to the particle position by
\begin{align}
V(t) &= p x(t)^2 + q x(t) + r \\
&= \frac{p X^2}{2} (1 - \cos{2\omega t}) + q X \sin{\omega t} + r
\end{align}
where $q$ and $p$ are conversion factors from particle displacement to SQUID voltage; $q$ describes the linear response of the SQUID, while $p$ describes the quadratic response, $r$ is an offset.
We see that
\begin{equation}
    \frac{p}{q} = \frac{u}{v}
\end{equation}

In the power spectrum of the voltage signal, the area of the fundamental peak is $A_\mathrm{f}$ and the area of the harmonic peak is $A_\mathrm{h}$.
And so
\begin{equation}
A_\mathrm{h} = \frac{p^2 X^4}{8}
\end{equation}
and
\begin{equation}
A_\mathrm{f} = \frac{q^2 X^2}{2}
\end{equation}
Earlier we defined
\begin{equation}
\eta A_\mathrm{f} = \langle x^2 \rangle = \frac{X^2}{2}
\end{equation}
So, we identify
\begin{equation}
\eta = \frac{1}{q^2}
\end{equation}

Overall, the nonlinearity of the pickup should give

\begin{equation}
R^\mathrm{PU} = \frac{A_\mathrm{h}}{A_\mathrm{f}^2} = \frac{u^2 \eta}{2 v^2}
\end{equation}
We extract $u_i$ and $v_i$ for the different directions $x$, $y$ and $z$ using the data in \cref{fig:pickupsignal}, and we extract $\eta$ from the comparison of the observed frequency pulling (see \cref{app:pulling}) with the trap anharmonicities given by FEM (see \cref{app:anharmonicity}).
And so, we estimate $R^\mathrm{PU}=\{2.5\cdot10^4,2.3\cdot10^4,1.6\cdot10^{-3}\}\,\mathrm{V^{-2}}$.
These are order-of-magnitude estimations, since the estimations of $v_i$ depend critically on our estimation of the position of the trap center.
The estimates $R^\mathrm{PU}$ are similar in magnitude to the observed values of $R^\mathrm{obs}$ for the x and y modes, and so we believe the spectral peaks at the second harmonics arise from the nonlinear pickup.

\section{Frequency pulling}\label{app:pulling}

\begin{figure*}[t!hbp]
    \centering
    \includegraphics[width=\textwidth]{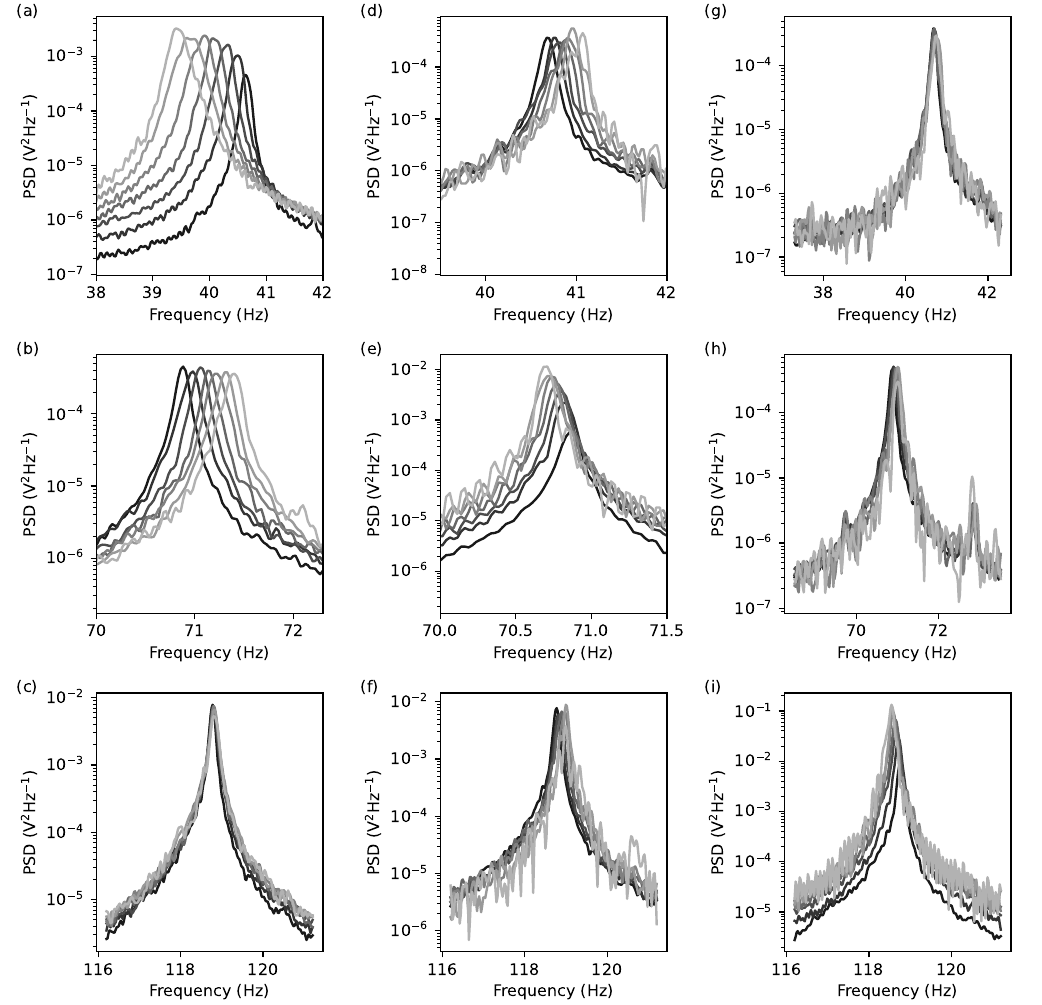}
    \caption{Frequency pulling. The spectral peaks corresponding to the COM frequencies depend on the COM mode amplitudes. In each panel, lighter colored spectra have higher mode amplitudes. In (a-c) $\langle x^2 \rangle$ is different for different spectra  (both $\langle y^2 \rangle$ and $\langle z^2 \rangle$ were relatively low). Similarly, in (d-f) $\langle y^2 \rangle$ is different for different spectra (both $\langle x^2 \rangle$ and $\langle z^2 \rangle$ were relatively low).
    Similarly, in (g-i) $\langle z^2 \rangle$ is different for different spectra (both $\langle x^2 \rangle$ and $\langle y^2 \rangle$ were relatively low).}
    \label{fig:quartic_PSDs}
\end{figure*}

\begin{figure*}[t!hbp]
    \centering
    \includegraphics[width=\textwidth]{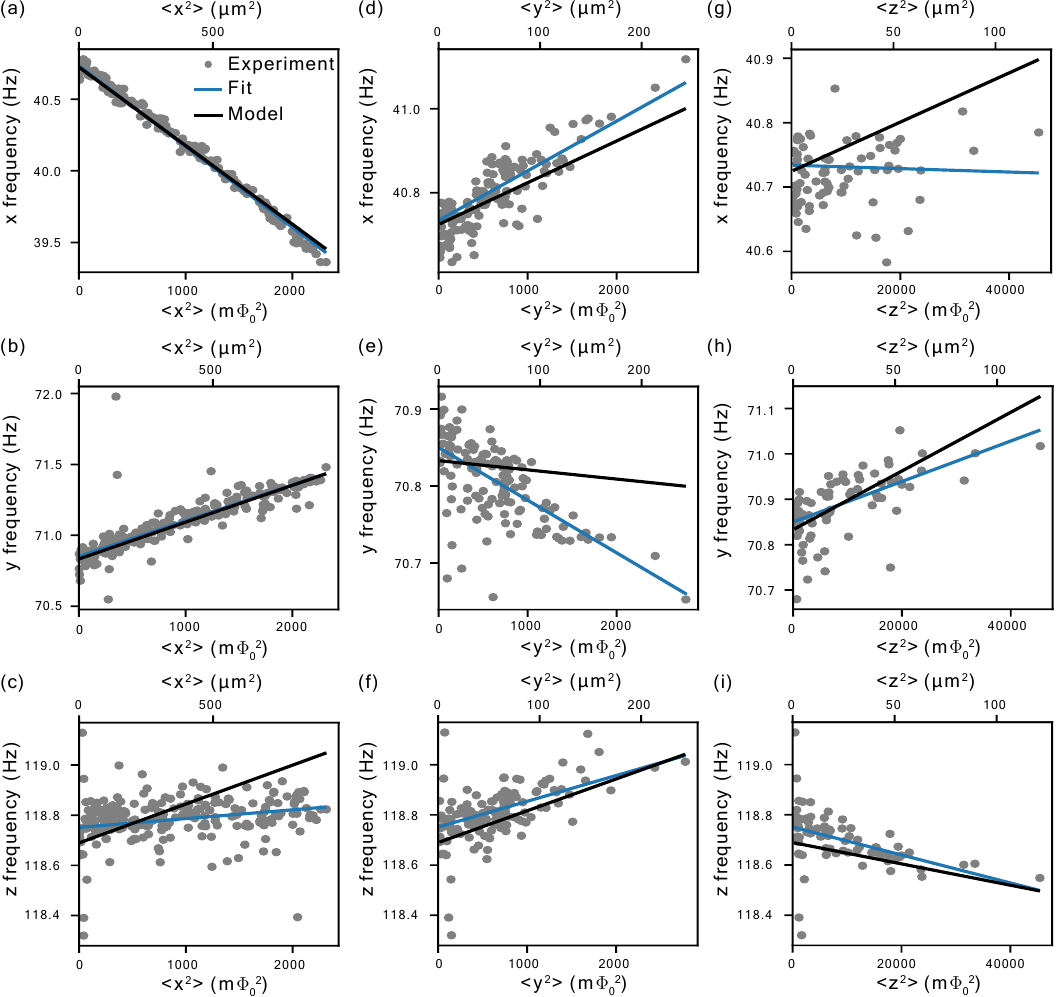}
    \caption{Frequency pulling. The COM frequencies vary depending on each of the mode amplitudes. Linear fits to the data are represented by blue lines, while the linear relations obtained using the trap anharmonicities extracted from FEM are shown by black lines. In most cases, both lines show a similar slope and capture the behavior reasonably well.}
    \label{fig:quartic_fits}
\end{figure*}

Figures \cref{fig:quartic_PSDs} and \cref{fig:quartic_fits} employ the same dataset and analysis method as was used to construct Fig.~4(a-d) of the main text.
While Fig.~4(a-d) shows the dependence of $\omega_x$ and $\omega_y$ on $\langle x^2 \rangle$, \cref{fig:quartic_fits} and \cref{fig:quartic_PSDs} show the dependence of $\omega_x$, $\omega_y$ and $\omega_z$ on $\langle x^2 \rangle$, $\langle y^2 \rangle$ and $\langle z^2 \rangle$.

In \cref{fig:quartic_PSDs} the blue-colored straight lines represent linear fits to the data, they are meant as guides to the eye.
The model curves are given by Eq.~(3) in the main text.
We extract estimates of the $\gamma_{ij}$ values from FEM (details in \cref{app:anharmonicity}).
In \cref{fig:quartic_fits}(a-i) the x-axes are the mean-square displacements in units of $\mathrm{m\upphi_0^2}$, and thus the gradients of the nine model lines depend on the three pickup efficiencies $\eta_i$ [In (a-c) the gradients depend on $\eta_x$, (d-f) depend on $\eta_y$ and (g-i) depend on $\eta_z$].
We treat the three pickup efficiencies as free parameters, and we obtain three estimates for each $\eta_i$, from the data in each of the panels. We choose the values which provide the best agreement between the observations and the model.
In this way we use the frequency pulling data to estimate the pickup efficiencies $\eta_i$.

\section{Frequency jumps}\label{app:frequencyjumps}

\begin{figure}[t!hbp]
    \centering
    \includegraphics{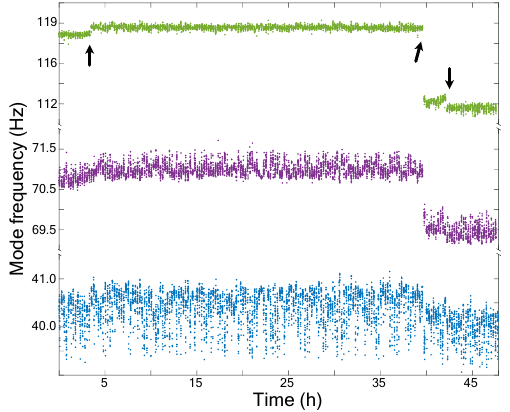}
    \caption{COM frequencies measured over 48\,h. The arrows indicate times when there were sudden simultaneous jumps of all the COM frequencies.}
    \label{fig:fluxjumps}
\end{figure}

When we levitate a particle we occasionally observe sudden changes of all the COM frequencies at once.
These changes appear to happen at random times.
We expect these frequency jumps are due to reorientation of trapped flux in the particle, or due to a change of the amount of the trapped flux in the particle.
In \cref{fig:fluxjumps} three such frequency jumps are indicated by arrows.
Here the particle was levitated for two days.
The 35\,h dataset used in Fig.~3 and Fig.~4 used the data between time 4\,h and 39\,h of this dataset.

\clearpage
\bibliography{bibliography}

\providecommand{\noopsort}[1]{}\providecommand{\singleletter}[1]{#1}
\begin{thebibliography}{76}%
\makeatletter
\providecommand \@ifxundefined [1]{%
 \@ifx{#1\undefined}
}%
\providecommand \@ifnum [1]{%
 \ifnum #1\expandafter \@firstoftwo
 \else \expandafter \@secondoftwo
 \fi
}%
\providecommand \@ifx [1]{%
 \ifx #1\expandafter \@firstoftwo
 \else \expandafter \@secondoftwo
 \fi
}%
\providecommand \natexlab [1]{#1}%
\providecommand \enquote  [1]{``#1''}%
\providecommand \bibnamefont  [1]{#1}%
\providecommand \bibfnamefont [1]{#1}%
\providecommand \citenamefont [1]{#1}%
\providecommand \href@noop [0]{\@secondoftwo}%
\providecommand \href [0]{\begingroup \@sanitize@url \@href}%
\providecommand \@href[1]{\@@startlink{#1}\@@href}%
\providecommand \@@href[1]{\endgroup#1\@@endlink}%
\providecommand \@sanitize@url [0]{\catcode `\\12\catcode `\$12\catcode
  `\&12\catcode `\#12\catcode `\^12\catcode `\_12\catcode `\%12\relax}%
\providecommand \@@startlink[1]{}%
\providecommand \@@endlink[0]{}%
\providecommand \url  [0]{\begingroup\@sanitize@url \@url }%
\providecommand \@url [1]{\endgroup\@href {#1}{\urlprefix }}%
\providecommand \urlprefix  [0]{URL }%
\providecommand \Eprint [0]{\href }%
\providecommand \doibase [0]{https://doi.org/}%
\providecommand \selectlanguage [0]{\@gobble}%
\providecommand \bibinfo  [0]{\@secondoftwo}%
\providecommand \bibfield  [0]{\@secondoftwo}%
\providecommand \translation [1]{[#1]}%
\providecommand \BibitemOpen [0]{}%
\providecommand \bibitemStop [0]{}%
\providecommand \bibitemNoStop [0]{.\EOS\space}%
\providecommand \EOS [0]{\spacefactor3000\relax}%
\providecommand \BibitemShut  [1]{\csname bibitem#1\endcsname}%
\let\auto@bib@innerbib\@empty
\bibitem [{\citenamefont {Ashkin}\ and\ \citenamefont
  {Dziedzic}(1976)}]{ashkinOpticalLevitationHigh1976}%
  \BibitemOpen
  \bibfield  {author} {\bibinfo {author} {\bibfnamefont {A.}~\bibnamefont
  {Ashkin}}\ and\ \bibinfo {author} {\bibfnamefont {J.~M.}\ \bibnamefont
  {Dziedzic}},\ }\bibfield  {title} {\bibinfo {title} {Optical levitation in
  high vacuum},\ }\href {https://doi.org/10.1063/1.88748} {\bibfield  {journal}
  {\bibinfo  {journal} {Appl. Phys. Lett.}\ }\textbf {\bibinfo {volume} {28}},\
  \bibinfo {pages} {333} (\bibinfo {year} {1976})}\BibitemShut {NoStop}%
\bibitem [{\citenamefont {Millen}\ \emph {et~al.}(2020)\citenamefont {Millen},
  \citenamefont {Monteiro}, \citenamefont {Pettit},\ and\ \citenamefont
  {Vamivakas}}]{millenOptomechanicsLevitatedParticles2020}%
  \BibitemOpen
  \bibfield  {author} {\bibinfo {author} {\bibfnamefont {J.}~\bibnamefont
  {Millen}}, \bibinfo {author} {\bibfnamefont {T.~S.}\ \bibnamefont
  {Monteiro}}, \bibinfo {author} {\bibfnamefont {R.}~\bibnamefont {Pettit}},\
  and\ \bibinfo {author} {\bibfnamefont {A.~N.}\ \bibnamefont {Vamivakas}},\
  }\bibfield  {title} {\bibinfo {title} {Optomechanics with levitated
  particles},\ }\href {https://doi.org/10.1088/1361-6633/ab6100} {\bibfield
  {journal} {\bibinfo  {journal} {Rep. Prog. Phys.}\ }\textbf {\bibinfo
  {volume} {83}},\ \bibinfo {pages} {026401} (\bibinfo {year}
  {2020})}\BibitemShut {NoStop}%
\bibitem [{\citenamefont {Gonzalez-Ballestero}\ \emph
  {et~al.}(2021)\citenamefont {Gonzalez-Ballestero}, \citenamefont
  {Aspelmeyer}, \citenamefont {Novotny}, \citenamefont {Quidant},\ and\
  \citenamefont {Romero-Isart}}]{levitodynamics}%
  \BibitemOpen
  \bibfield  {author} {\bibinfo {author} {\bibfnamefont {C.}~\bibnamefont
  {Gonzalez-Ballestero}}, \bibinfo {author} {\bibfnamefont {M.}~\bibnamefont
  {Aspelmeyer}}, \bibinfo {author} {\bibfnamefont {L.}~\bibnamefont {Novotny}},
  \bibinfo {author} {\bibfnamefont {R.}~\bibnamefont {Quidant}},\ and\ \bibinfo
  {author} {\bibfnamefont {O.}~\bibnamefont {Romero-Isart}},\ }\bibfield
  {title} {\bibinfo {title} {Levitodynamics: Levitation and control of
  microscopic objects in vacuum},\ }\href
  {https://doi.org/10.1126/science.abg3027} {\bibfield  {journal} {\bibinfo
  {journal} {Science}\ }\textbf {\bibinfo {volume} {374}},\ \bibinfo {pages}
  {3027} (\bibinfo {year} {2021})}\BibitemShut {NoStop}%
\bibitem [{\citenamefont {Ranjit}\ \emph {et~al.}(2016)\citenamefont {Ranjit},
  \citenamefont {Cunningham}, \citenamefont {Casey},\ and\ \citenamefont
  {Geraci}}]{Gambhir_2016}%
  \BibitemOpen
  \bibfield  {author} {\bibinfo {author} {\bibfnamefont {G.}~\bibnamefont
  {Ranjit}}, \bibinfo {author} {\bibfnamefont {M.}~\bibnamefont {Cunningham}},
  \bibinfo {author} {\bibfnamefont {K.}~\bibnamefont {Casey}},\ and\ \bibinfo
  {author} {\bibfnamefont {A.~A.}\ \bibnamefont {Geraci}},\ }\bibfield  {title}
  {\bibinfo {title} {Zeptonewton force sensing with nanospheres in an optical
  lattice},\ }\href {https://doi.org/10.1103/PhysRevA.93.053801} {\bibfield
  {journal} {\bibinfo  {journal} {Phys. Rev. A}\ }\textbf {\bibinfo {volume}
  {93}},\ \bibinfo {pages} {053801} (\bibinfo {year} {2016})}\BibitemShut
  {NoStop}%
\bibitem [{\citenamefont {Prat-Camps}\ \emph {et~al.}(2017)\citenamefont
  {Prat-Camps}, \citenamefont {Teo}, \citenamefont {Rusconi}, \citenamefont
  {Wieczorek},\ and\ \citenamefont {Romero-Isart}}]{Prat-Camps_2017}%
  \BibitemOpen
  \bibfield  {author} {\bibinfo {author} {\bibfnamefont {J.}~\bibnamefont
  {Prat-Camps}}, \bibinfo {author} {\bibfnamefont {C.}~\bibnamefont {Teo}},
  \bibinfo {author} {\bibfnamefont {C.~C.}\ \bibnamefont {Rusconi}}, \bibinfo
  {author} {\bibfnamefont {W.}~\bibnamefont {Wieczorek}},\ and\ \bibinfo
  {author} {\bibfnamefont {O.}~\bibnamefont {Romero-Isart}},\ }\bibfield
  {title} {\bibinfo {title} {Ultrasensitive inertial and force sensors with
  diamagnetically levitated magnets},\ }\href
  {https://doi.org/10.1103/PhysRevApplied.8.034002} {\bibfield  {journal}
  {\bibinfo  {journal} {Phys. Rev. Appl.}\ }\textbf {\bibinfo {volume} {8}},\
  \bibinfo {pages} {034002} (\bibinfo {year} {2017})}\BibitemShut {NoStop}%
\bibitem [{\citenamefont {Monteiro}\ \emph {et~al.}(2020)\citenamefont
  {Monteiro}, \citenamefont {Li}, \citenamefont {Afek}, \citenamefont {Li},
  \citenamefont {Mossman},\ and\ \citenamefont
  {Moore}}]{monteiroForceAccelerationSensing2020}%
  \BibitemOpen
  \bibfield  {author} {\bibinfo {author} {\bibfnamefont {F.}~\bibnamefont
  {Monteiro}}, \bibinfo {author} {\bibfnamefont {W.}~\bibnamefont {Li}},
  \bibinfo {author} {\bibfnamefont {G.}~\bibnamefont {Afek}}, \bibinfo {author}
  {\bibfnamefont {C.-l.}\ \bibnamefont {Li}}, \bibinfo {author} {\bibfnamefont
  {M.}~\bibnamefont {Mossman}},\ and\ \bibinfo {author} {\bibfnamefont {D.~C.}\
  \bibnamefont {Moore}},\ }\bibfield  {title} {\bibinfo {title} {{Force and
  Acceleration Sensing with Optically Levitated Nanogram Masses at Microkelvin
  Temperatures}},\ }\href {https://doi.org/10.1103/PhysRevA.101.053835}
  {\bibfield  {journal} {\bibinfo  {journal} {Phys. Rev. A}\ }\textbf {\bibinfo
  {volume} {101}},\ \bibinfo {pages} {053835} (\bibinfo {year}
  {2020})}\BibitemShut {NoStop}%
\bibitem [{\citenamefont {Weiss}\ \emph {et~al.}(2021)\citenamefont {Weiss},
  \citenamefont {Roda-Llordes}, \citenamefont {Torrontegui}, \citenamefont
  {Aspelmeyer},\ and\ \citenamefont {Romero-Isart}}]{Weiss2021}%
  \BibitemOpen
  \bibfield  {author} {\bibinfo {author} {\bibfnamefont {T.}~\bibnamefont
  {Weiss}}, \bibinfo {author} {\bibfnamefont {M.}~\bibnamefont {Roda-Llordes}},
  \bibinfo {author} {\bibfnamefont {E.}~\bibnamefont {Torrontegui}}, \bibinfo
  {author} {\bibfnamefont {M.}~\bibnamefont {Aspelmeyer}},\ and\ \bibinfo
  {author} {\bibfnamefont {O.}~\bibnamefont {Romero-Isart}},\ }\bibfield
  {title} {\bibinfo {title} {Large quantum delocalization of a levitated
  nanoparticle using optimal control: Applications for force sensing and
  entangling via weak forces},\ }\href
  {https://doi.org/10.1103/PHYSREVLETT.127.023601} {\bibfield  {journal}
  {\bibinfo  {journal} {Phys. Rev. Lett.}\ }\textbf {\bibinfo {volume} {127}},\
  \bibinfo {pages} {023601} (\bibinfo {year} {2021})}\BibitemShut {NoStop}%
\bibitem [{\citenamefont {Goodkind}(1999)}]{Goodkind1999}%
  \BibitemOpen
  \bibfield  {author} {\bibinfo {author} {\bibfnamefont {J.~M.}\ \bibnamefont
  {Goodkind}},\ }\bibfield  {title} {\bibinfo {title} {The superconducting
  gravimeter},\ }\href {https://doi.org/10.1063/1.1150092} {\bibfield
  {journal} {\bibinfo  {journal} {Rev. Sci. Instrum.}\ }\textbf {\bibinfo
  {volume} {70}},\ \bibinfo {pages} {4131} (\bibinfo {year}
  {1999})}\BibitemShut {NoStop}%
\bibitem [{\citenamefont {Moody}\ \emph {et~al.}(2002)\citenamefont {Moody},
  \citenamefont {Paik},\ and\ \citenamefont {Canavan}}]{canavan_2002}%
  \BibitemOpen
  \bibfield  {author} {\bibinfo {author} {\bibfnamefont {M.~V.}\ \bibnamefont
  {Moody}}, \bibinfo {author} {\bibfnamefont {H.~J.}\ \bibnamefont {Paik}},\
  and\ \bibinfo {author} {\bibfnamefont {E.~R.}\ \bibnamefont {Canavan}},\
  }\bibfield  {title} {\bibinfo {title} {Three-axis superconducting gravity
  gradiometer for sensitive gravity experiments},\ }\href
  {https://doi.org/10.1063/1.1511798} {\bibfield  {journal} {\bibinfo
  {journal} {Rev. Sci. Instrum.}\ }\textbf {\bibinfo {volume} {73}},\ \bibinfo
  {pages} {3957} (\bibinfo {year} {2002})}\BibitemShut {NoStop}%
\bibitem [{\citenamefont {Johnsson}\ \emph {et~al.}(2016)\citenamefont
  {Johnsson}, \citenamefont {Brennen},\ and\ \citenamefont
  {Twamley}}]{Johnsson_2016}%
  \BibitemOpen
  \bibfield  {author} {\bibinfo {author} {\bibfnamefont {M.~T.}\ \bibnamefont
  {Johnsson}}, \bibinfo {author} {\bibfnamefont {G.~K.}\ \bibnamefont
  {Brennen}},\ and\ \bibinfo {author} {\bibfnamefont {J.}~\bibnamefont
  {Twamley}},\ }\bibfield  {title} {\bibinfo {title} {Macroscopic
  superpositions and gravimetry with quantum magnetomechanics},\ }\href
  {https://doi.org/10.1038/srep37495} {\bibfield  {journal} {\bibinfo
  {journal} {Sci. Rep.}\ }\textbf {\bibinfo {volume} {6}},\ \bibinfo {pages}
  {37495} (\bibinfo {year} {2016})}\BibitemShut {NoStop}%
\bibitem [{\citenamefont {Monteiro}\ \emph {et~al.}(2017)\citenamefont
  {Monteiro}, \citenamefont {Ghosh}, \citenamefont {Fine},\ and\ \citenamefont
  {Moore}}]{monteiroOpticalLevitation10ng2017}%
  \BibitemOpen
  \bibfield  {author} {\bibinfo {author} {\bibfnamefont {F.}~\bibnamefont
  {Monteiro}}, \bibinfo {author} {\bibfnamefont {S.}~\bibnamefont {Ghosh}},
  \bibinfo {author} {\bibfnamefont {A.~G.}\ \bibnamefont {Fine}},\ and\
  \bibinfo {author} {\bibfnamefont {D.~C.}\ \bibnamefont {Moore}},\ }\bibfield
  {title} {\bibinfo {title} {Optical levitation of 10-ng spheres with nano-g
  acceleration sensitivity},\ }\href
  {https://doi.org/10.1103/PhysRevA.96.063841} {\bibfield  {journal} {\bibinfo
  {journal} {Phys. Rev. A}\ }\textbf {\bibinfo {volume} {96}},\ \bibinfo
  {pages} {063841} (\bibinfo {year} {2017})}\BibitemShut {NoStop}%
\bibitem [{\citenamefont {Timberlake}\ \emph {et~al.}(2019)\citenamefont
  {Timberlake}, \citenamefont {Gasbarri}, \citenamefont {Vinante},
  \citenamefont {Setter},\ and\ \citenamefont
  {Ulbricht}}]{timberlake_acceleration_2019}%
  \BibitemOpen
  \bibfield  {author} {\bibinfo {author} {\bibfnamefont {C.}~\bibnamefont
  {Timberlake}}, \bibinfo {author} {\bibfnamefont {G.}~\bibnamefont
  {Gasbarri}}, \bibinfo {author} {\bibfnamefont {A.}~\bibnamefont {Vinante}},
  \bibinfo {author} {\bibfnamefont {A.}~\bibnamefont {Setter}},\ and\ \bibinfo
  {author} {\bibfnamefont {H.}~\bibnamefont {Ulbricht}},\ }\bibfield  {title}
  {\bibinfo {title} {Acceleration sensing with magnetically levitated
  oscillators above a superconductor},\ }\href
  {https://doi.org/10.1063/1.5129145} {\bibfield  {journal} {\bibinfo
  {journal} {Appl. Phys. Lett.}\ }\textbf {\bibinfo {volume} {115}},\ \bibinfo
  {pages} {224101} (\bibinfo {year} {2019})}\BibitemShut {NoStop}%
\bibitem [{\citenamefont {Li}\ \emph {et~al.}(2010)\citenamefont {Li},
  \citenamefont {Kheifets}, \citenamefont {Medellin},\ and\ \citenamefont
  {Raizen}}]{Li2010}%
  \BibitemOpen
  \bibfield  {author} {\bibinfo {author} {\bibfnamefont {T.}~\bibnamefont
  {Li}}, \bibinfo {author} {\bibfnamefont {S.}~\bibnamefont {Kheifets}},
  \bibinfo {author} {\bibfnamefont {D.}~\bibnamefont {Medellin}},\ and\
  \bibinfo {author} {\bibfnamefont {M.~G.}\ \bibnamefont {Raizen}},\ }\bibfield
   {title} {\bibinfo {title} {Measurement of the instantaneous velocity of a
  brownian particle},\ }\href {https://doi.org/10.1126/SCIENCE.1189403}
  {\bibfield  {journal} {\bibinfo  {journal} {Science}\ }\textbf {\bibinfo
  {volume} {328}},\ \bibinfo {pages} {1673} (\bibinfo {year}
  {2010})}\BibitemShut {NoStop}%
\bibitem [{\citenamefont {Gieseler}\ and\ \citenamefont
  {Millen}(2018)}]{gieselerLevitatedNanoparticlesMicroscopic2018}%
  \BibitemOpen
  \bibfield  {author} {\bibinfo {author} {\bibfnamefont {J.}~\bibnamefont
  {Gieseler}}\ and\ \bibinfo {author} {\bibfnamefont {J.}~\bibnamefont
  {Millen}},\ }\bibfield  {title} {\bibinfo {title} {Levitated
  {{Nanoparticles}} for {{Microscopic Thermodynamics}}\textemdash{{A
  Review}}},\ }\href {https://doi.org/10.3390/e20050326} {\bibfield  {journal}
  {\bibinfo  {journal} {Entropy}\ }\textbf {\bibinfo {volume} {20}},\ \bibinfo
  {pages} {326} (\bibinfo {year} {2018})}\BibitemShut {NoStop}%
\bibitem [{\citenamefont {Debiossac}\ \emph {et~al.}(2020)\citenamefont
  {Debiossac}, \citenamefont {Grass}, \citenamefont {Alonso}, \citenamefont
  {Lutz},\ and\ \citenamefont
  {Kiesel}}]{debiossacThermodynamicsContinuousNonMarkovian2020}%
  \BibitemOpen
  \bibfield  {author} {\bibinfo {author} {\bibfnamefont {M.}~\bibnamefont
  {Debiossac}}, \bibinfo {author} {\bibfnamefont {D.}~\bibnamefont {Grass}},
  \bibinfo {author} {\bibfnamefont {J.~J.}\ \bibnamefont {Alonso}}, \bibinfo
  {author} {\bibfnamefont {E.}~\bibnamefont {Lutz}},\ and\ \bibinfo {author}
  {\bibfnamefont {N.}~\bibnamefont {Kiesel}},\ }\bibfield  {title} {\bibinfo
  {title} {Thermodynamics of continuous non-{{Markovian}} feedback control},\
  }\href {https://doi.org/10.1038/s41467-020-15148-5} {\bibfield  {journal}
  {\bibinfo  {journal} {Nat. Commun.}\ }\textbf {\bibinfo {volume} {11}},\
  \bibinfo {pages} {1360} (\bibinfo {year} {2020})}\BibitemShut {NoStop}%
\bibitem [{\citenamefont {Dholakia}\ and\ \citenamefont
  {Zem\'anek}(2010)}]{Dholakia2010}%
  \BibitemOpen
  \bibfield  {author} {\bibinfo {author} {\bibfnamefont {K.}~\bibnamefont
  {Dholakia}}\ and\ \bibinfo {author} {\bibfnamefont {P.}~\bibnamefont
  {Zem\'anek}},\ }\bibfield  {title} {\bibinfo {title} {Colloquium: Gripped by
  light: Optical binding},\ }\href {https://doi.org/10.1103/RevModPhys.82.1767}
  {\bibfield  {journal} {\bibinfo  {journal} {Rev. Mod. Phys.}\ }\textbf
  {\bibinfo {volume} {82}},\ \bibinfo {pages} {1767} (\bibinfo {year}
  {2010})}\BibitemShut {NoStop}%
\bibitem [{\citenamefont {Lechner}\ \emph {et~al.}(2013)\citenamefont
  {Lechner}, \citenamefont {Habraken}, \citenamefont {Kiesel}, \citenamefont
  {Aspelmeyer},\ and\ \citenamefont {Zoller}}]{Lechner2013}%
  \BibitemOpen
  \bibfield  {author} {\bibinfo {author} {\bibfnamefont {W.}~\bibnamefont
  {Lechner}}, \bibinfo {author} {\bibfnamefont {S.~J.~M.}\ \bibnamefont
  {Habraken}}, \bibinfo {author} {\bibfnamefont {N.}~\bibnamefont {Kiesel}},
  \bibinfo {author} {\bibfnamefont {M.}~\bibnamefont {Aspelmeyer}},\ and\
  \bibinfo {author} {\bibfnamefont {P.}~\bibnamefont {Zoller}},\ }\bibfield
  {title} {\bibinfo {title} {Cavity optomechanics of levitated nanodumbbells:
  Nonequilibrium phases and self-assembly},\ }\href
  {https://doi.org/10.1103/PhysRevLett.110.143604} {\bibfield  {journal}
  {\bibinfo  {journal} {Phys. Rev. Lett.}\ }\textbf {\bibinfo {volume} {110}},\
  \bibinfo {pages} {143604} (\bibinfo {year} {2013})}\BibitemShut {NoStop}%
\bibitem [{\citenamefont {Simpson}\ \emph {et~al.}(2016)\citenamefont
  {Simpson}, \citenamefont {Chv\'atal},\ and\ \citenamefont
  {Zem\'anek}}]{Simpson2016}%
  \BibitemOpen
  \bibfield  {author} {\bibinfo {author} {\bibfnamefont {S.~H.}\ \bibnamefont
  {Simpson}}, \bibinfo {author} {\bibfnamefont {L.}~\bibnamefont {Chv\'atal}},\
  and\ \bibinfo {author} {\bibfnamefont {P.}~\bibnamefont {Zem\'anek}},\
  }\bibfield  {title} {\bibinfo {title} {Synchronization of colloidal rotors
  through angular optical binding},\ }\href
  {https://doi.org/10.1103/PhysRevA.93.023842} {\bibfield  {journal} {\bibinfo
  {journal} {Phys. Rev. A}\ }\textbf {\bibinfo {volume} {93}},\ \bibinfo
  {pages} {023842} (\bibinfo {year} {2016})}\BibitemShut {NoStop}%
\bibitem [{\citenamefont {Yan}\ \emph {et~al.}(2022)\citenamefont {Yan},
  \citenamefont {Yu}, \citenamefont {Han}, \citenamefont {Li},\ and\
  \citenamefont {Zhang}}]{Yan2022}%
  \BibitemOpen
  \bibfield  {author} {\bibinfo {author} {\bibfnamefont {J.}~\bibnamefont
  {Yan}}, \bibinfo {author} {\bibfnamefont {X.}~\bibnamefont {Yu}}, \bibinfo
  {author} {\bibfnamefont {Z.~V.}\ \bibnamefont {Han}}, \bibinfo {author}
  {\bibfnamefont {T.}~\bibnamefont {Li}},\ and\ \bibinfo {author}
  {\bibfnamefont {J.}~\bibnamefont {Zhang}},\ }\href
  {https://doi.org/10.48550/arXiv.2207.03641} {\bibinfo {title} {On-demand
  assembly of optically-levitated nanoparticle arrays in vacuum}} (\bibinfo
  {year} {2022}),\ \Eprint {https://arxiv.org/abs/arXiv:2207.03641}
  {arXiv:2207.03641} \BibitemShut {NoStop}%
\bibitem [{\citenamefont {Rieser}\ \emph {et~al.}(2022)\citenamefont {Rieser},
  \citenamefont {Ciampini}, \citenamefont {Rudolph}, \citenamefont {Kiesel},
  \citenamefont {Hornberger}, \citenamefont {Stickler}, \citenamefont
  {Aspelmeyer},\ and\ \citenamefont {Delić}}]{Rieser2022}%
  \BibitemOpen
  \bibfield  {author} {\bibinfo {author} {\bibfnamefont {J.}~\bibnamefont
  {Rieser}}, \bibinfo {author} {\bibfnamefont {M.~A.}\ \bibnamefont
  {Ciampini}}, \bibinfo {author} {\bibfnamefont {H.}~\bibnamefont {Rudolph}},
  \bibinfo {author} {\bibfnamefont {N.}~\bibnamefont {Kiesel}}, \bibinfo
  {author} {\bibfnamefont {K.}~\bibnamefont {Hornberger}}, \bibinfo {author}
  {\bibfnamefont {B.~A.}\ \bibnamefont {Stickler}}, \bibinfo {author}
  {\bibfnamefont {M.}~\bibnamefont {Aspelmeyer}},\ and\ \bibinfo {author}
  {\bibfnamefont {U.}~\bibnamefont {Delić}},\ }\bibfield  {title} {\bibinfo
  {title} {Tunable light-induced dipole-dipole interaction between optically
  levitated nanoparticles},\ }\href {https://doi.org/10.1126/science.abp9941}
  {\bibfield  {journal} {\bibinfo  {journal} {Science}\ }\textbf {\bibinfo
  {volume} {377}},\ \bibinfo {pages} {987} (\bibinfo {year}
  {2022})}\BibitemShut {NoStop}%
\bibitem [{\citenamefont {Delord}\ \emph {et~al.}(2017)\citenamefont {Delord},
  \citenamefont {Nicolas}, \citenamefont {Schwab},\ and\ \citenamefont
  {Hétet}}]{Delord_2017}%
  \BibitemOpen
  \bibfield  {author} {\bibinfo {author} {\bibfnamefont {T.}~\bibnamefont
  {Delord}}, \bibinfo {author} {\bibfnamefont {L.}~\bibnamefont {Nicolas}},
  \bibinfo {author} {\bibfnamefont {L.}~\bibnamefont {Schwab}},\ and\ \bibinfo
  {author} {\bibfnamefont {G.}~\bibnamefont {Hétet}},\ }\bibfield  {title}
  {\bibinfo {title} {Electron spin resonance from nv centers in diamonds
  levitating in an ion trap},\ }\href
  {https://doi.org/10.1088/1367-2630/aa659c} {\bibfield  {journal} {\bibinfo
  {journal} {New J. Phys.}\ }\textbf {\bibinfo {volume} {19}},\ \bibinfo
  {pages} {033031} (\bibinfo {year} {2017})}\BibitemShut {NoStop}%
\bibitem [{\citenamefont {Perdriat}\ \emph {et~al.}(2021)\citenamefont
  {Perdriat}, \citenamefont {Pellet-Mary}, \citenamefont {Huillery},
  \citenamefont {Rondin},\ and\ \citenamefont {Hétet}}]{Pedriat_2021}%
  \BibitemOpen
  \bibfield  {author} {\bibinfo {author} {\bibfnamefont {M.}~\bibnamefont
  {Perdriat}}, \bibinfo {author} {\bibfnamefont {C.}~\bibnamefont
  {Pellet-Mary}}, \bibinfo {author} {\bibfnamefont {P.}~\bibnamefont
  {Huillery}}, \bibinfo {author} {\bibfnamefont {L.}~\bibnamefont {Rondin}},\
  and\ \bibinfo {author} {\bibfnamefont {G.}~\bibnamefont {Hétet}},\
  }\bibfield  {title} {\bibinfo {title} {Spin-mechanics with nitrogen-vacancy
  centers and trapped particles},\ }\href {https://doi.org/10.3390/mi12060651}
  {\bibfield  {journal} {\bibinfo  {journal} {Micromachines}\ }\textbf
  {\bibinfo {volume} {12}},\ \bibinfo {pages} {651} (\bibinfo {year}
  {2021})}\BibitemShut {NoStop}%
\bibitem [{\citenamefont {Stickler}\ \emph {et~al.}(2021)\citenamefont
  {Stickler}, \citenamefont {Hornberger},\ and\ \citenamefont
  {Kim}}]{Stickler2021}%
  \BibitemOpen
  \bibfield  {author} {\bibinfo {author} {\bibfnamefont {B.~A.}\ \bibnamefont
  {Stickler}}, \bibinfo {author} {\bibfnamefont {K.}~\bibnamefont
  {Hornberger}},\ and\ \bibinfo {author} {\bibfnamefont {M.~S.}\ \bibnamefont
  {Kim}},\ }\bibfield  {title} {\bibinfo {title} {Quantum rotations of
  nanoparticles},\ }\href {https://doi.org/10.1038/s42254-021-00335-0}
  {\bibfield  {journal} {\bibinfo  {journal} {Nat. Rev. Phys.}\ }\textbf
  {\bibinfo {volume} {3}},\ \bibinfo {pages} {589} (\bibinfo {year}
  {2021})}\BibitemShut {NoStop}%
\bibitem [{\citenamefont {Rusconi}\ \emph {et~al.}(2022)\citenamefont
  {Rusconi}, \citenamefont {Perdriat}, \citenamefont {H\'etet}, \citenamefont
  {Romero-Isart},\ and\ \citenamefont {Stickler}}]{Cosimo_2022}%
  \BibitemOpen
  \bibfield  {author} {\bibinfo {author} {\bibfnamefont {C.~C.}\ \bibnamefont
  {Rusconi}}, \bibinfo {author} {\bibfnamefont {M.}~\bibnamefont {Perdriat}},
  \bibinfo {author} {\bibfnamefont {G.}~\bibnamefont {H\'etet}}, \bibinfo
  {author} {\bibfnamefont {O.}~\bibnamefont {Romero-Isart}},\ and\ \bibinfo
  {author} {\bibfnamefont {B.~A.}\ \bibnamefont {Stickler}},\ }\bibfield
  {title} {\bibinfo {title} {Spin-controlled quantum interference of levitated
  nanorotors},\ }\href {https://doi.org/10.1103/PhysRevLett.129.093605}
  {\bibfield  {journal} {\bibinfo  {journal} {Phys. Rev. Lett.}\ }\textbf
  {\bibinfo {volume} {129}},\ \bibinfo {pages} {093605} (\bibinfo {year}
  {2022})}\BibitemShut {NoStop}%
\bibitem [{\citenamefont {Delić}\ \emph {et~al.}(2020)\citenamefont {Delić},
  \citenamefont {Reisenbauer}, \citenamefont {Dare}, \citenamefont {Grass},
  \citenamefont {Vuletić}, \citenamefont {Kiesel},\ and\ \citenamefont
  {Aspelmeyer}}]{delic_cooling_2020}%
  \BibitemOpen
  \bibfield  {author} {\bibinfo {author} {\bibfnamefont {U.}~\bibnamefont
  {Delić}}, \bibinfo {author} {\bibfnamefont {M.}~\bibnamefont {Reisenbauer}},
  \bibinfo {author} {\bibfnamefont {K.}~\bibnamefont {Dare}}, \bibinfo {author}
  {\bibfnamefont {D.}~\bibnamefont {Grass}}, \bibinfo {author} {\bibfnamefont
  {V.}~\bibnamefont {Vuletić}}, \bibinfo {author} {\bibfnamefont
  {N.}~\bibnamefont {Kiesel}},\ and\ \bibinfo {author} {\bibfnamefont
  {M.}~\bibnamefont {Aspelmeyer}},\ }\bibfield  {title} {\bibinfo {title}
  {Cooling of a levitated nanoparticle to the motional quantum ground state},\
  }\href {https://doi.org/10.1126/science.aba3993} {\bibfield  {journal}
  {\bibinfo  {journal} {Science}\ }\textbf {\bibinfo {volume} {367}},\ \bibinfo
  {pages} {892} (\bibinfo {year} {2020})}\BibitemShut {NoStop}%
\bibitem [{\citenamefont {Magrini}\ \emph {et~al.}(2021)\citenamefont
  {Magrini}, \citenamefont {Rosenzweig}, \citenamefont {Bach}, \citenamefont
  {{Deutschmann-Olek}}, \citenamefont {Hofer}, \citenamefont {Hong},
  \citenamefont {Kiesel}, \citenamefont {Kugi},\ and\ \citenamefont
  {Aspelmeyer}}]{magriniRealtimeOptimalQuantum2021}%
  \BibitemOpen
  \bibfield  {author} {\bibinfo {author} {\bibfnamefont {L.}~\bibnamefont
  {Magrini}}, \bibinfo {author} {\bibfnamefont {P.}~\bibnamefont {Rosenzweig}},
  \bibinfo {author} {\bibfnamefont {C.}~\bibnamefont {Bach}}, \bibinfo {author}
  {\bibfnamefont {A.}~\bibnamefont {{Deutschmann-Olek}}}, \bibinfo {author}
  {\bibfnamefont {S.~G.}\ \bibnamefont {Hofer}}, \bibinfo {author}
  {\bibfnamefont {S.}~\bibnamefont {Hong}}, \bibinfo {author} {\bibfnamefont
  {N.}~\bibnamefont {Kiesel}}, \bibinfo {author} {\bibfnamefont
  {A.}~\bibnamefont {Kugi}},\ and\ \bibinfo {author} {\bibfnamefont
  {M.}~\bibnamefont {Aspelmeyer}},\ }\bibfield  {title} {\bibinfo {title}
  {Real-time optimal quantum control of mechanical motion at room
  temperature},\ }\href {https://doi.org/10.1038/s41586-021-03602-3} {\bibfield
   {journal} {\bibinfo  {journal} {Nature}\ }\textbf {\bibinfo {volume}
  {595}},\ \bibinfo {pages} {373} (\bibinfo {year} {2021})}\BibitemShut
  {NoStop}%
\bibitem [{\citenamefont {Tebbenjohanns}\ \emph {et~al.}(2021)\citenamefont
  {Tebbenjohanns}, \citenamefont {Mattana}, \citenamefont {Rossi},
  \citenamefont {Frimmer},\ and\ \citenamefont
  {Novotny}}]{tebbenjohannsQuantumControlNanoparticle2021}%
  \BibitemOpen
  \bibfield  {author} {\bibinfo {author} {\bibfnamefont {F.}~\bibnamefont
  {Tebbenjohanns}}, \bibinfo {author} {\bibfnamefont {M.~L.}\ \bibnamefont
  {Mattana}}, \bibinfo {author} {\bibfnamefont {M.}~\bibnamefont {Rossi}},
  \bibinfo {author} {\bibfnamefont {M.}~\bibnamefont {Frimmer}},\ and\ \bibinfo
  {author} {\bibfnamefont {L.}~\bibnamefont {Novotny}},\ }\bibfield  {title}
  {\bibinfo {title} {Quantum control of a nanoparticle optically levitated in
  cryogenic free space},\ }\href {https://doi.org/10.1038/s41586-021-03617-w}
  {\bibfield  {journal} {\bibinfo  {journal} {Nature}\ }\textbf {\bibinfo
  {volume} {595}},\ \bibinfo {pages} {378} (\bibinfo {year}
  {2021})}\BibitemShut {NoStop}%
\bibitem [{\citenamefont {Ranfagni}\ \emph {et~al.}(2022)\citenamefont
  {Ranfagni}, \citenamefont {B\o{}rkje}, \citenamefont {Marino},\ and\
  \citenamefont {Marin}}]{Ranfagni2022}%
  \BibitemOpen
  \bibfield  {author} {\bibinfo {author} {\bibfnamefont {A.}~\bibnamefont
  {Ranfagni}}, \bibinfo {author} {\bibfnamefont {K.}~\bibnamefont {B\o{}rkje}},
  \bibinfo {author} {\bibfnamefont {F.}~\bibnamefont {Marino}},\ and\ \bibinfo
  {author} {\bibfnamefont {F.}~\bibnamefont {Marin}},\ }\bibfield  {title}
  {\bibinfo {title} {Two-dimensional quantum motion of a levitated
  nanosphere},\ }\href {https://doi.org/10.1103/PhysRevResearch.4.033051}
  {\bibfield  {journal} {\bibinfo  {journal} {Phys. Rev. Research}\ }\textbf
  {\bibinfo {volume} {4}},\ \bibinfo {pages} {033051} (\bibinfo {year}
  {2022})}\BibitemShut {NoStop}%
\bibitem [{\citenamefont {Piotrowski}\ \emph {et~al.}(2023)\citenamefont
  {Piotrowski}, \citenamefont {Windey}, \citenamefont {Vijayan}, \citenamefont
  {{Gonzalez-Ballestero}}, \citenamefont {{de los R{\'i}os Sommer}},
  \citenamefont {Meyer}, \citenamefont {Quidant}, \citenamefont
  {{Romero-Isart}}, \citenamefont {Reimann},\ and\ \citenamefont
  {Novotny}}]{Piotrowski2023}%
  \BibitemOpen
  \bibfield  {author} {\bibinfo {author} {\bibfnamefont {J.}~\bibnamefont
  {Piotrowski}}, \bibinfo {author} {\bibfnamefont {D.}~\bibnamefont {Windey}},
  \bibinfo {author} {\bibfnamefont {J.}~\bibnamefont {Vijayan}}, \bibinfo
  {author} {\bibfnamefont {C.}~\bibnamefont {{Gonzalez-Ballestero}}}, \bibinfo
  {author} {\bibfnamefont {A.}~\bibnamefont {{de los R{\'i}os Sommer}}},
  \bibinfo {author} {\bibfnamefont {N.}~\bibnamefont {Meyer}}, \bibinfo
  {author} {\bibfnamefont {R.}~\bibnamefont {Quidant}}, \bibinfo {author}
  {\bibfnamefont {O.}~\bibnamefont {{Romero-Isart}}}, \bibinfo {author}
  {\bibfnamefont {R.}~\bibnamefont {Reimann}},\ and\ \bibinfo {author}
  {\bibfnamefont {L.}~\bibnamefont {Novotny}},\ }\bibfield  {title} {\bibinfo
  {title} {Simultaneous ground-state cooling of two mechanical modes of a
  levitated nanoparticle},\ }\href {https://doi.org/10.1038/s41567-023-01956-1}
  {\bibfield  {journal} {\bibinfo  {journal} {Nat. Phys.}\ ,\ \bibinfo {pages}
  {1}} (\bibinfo {year} {2023})}\BibitemShut {NoStop}%
\bibitem [{\citenamefont {Romero-Isart}\ \emph {et~al.}(2011)\citenamefont
  {Romero-Isart}, \citenamefont {Pflanzer}, \citenamefont {Blaser},
  \citenamefont {Kaltenbaek}, \citenamefont {Kiesel}, \citenamefont
  {Aspelmeyer},\ and\ \citenamefont
  {Cirac}}]{romero-isartLargeQuantumSuperpositions2011}%
  \BibitemOpen
  \bibfield  {author} {\bibinfo {author} {\bibfnamefont {O.}~\bibnamefont
  {Romero-Isart}}, \bibinfo {author} {\bibfnamefont {A.~C.}\ \bibnamefont
  {Pflanzer}}, \bibinfo {author} {\bibfnamefont {F.}~\bibnamefont {Blaser}},
  \bibinfo {author} {\bibfnamefont {R.}~\bibnamefont {Kaltenbaek}}, \bibinfo
  {author} {\bibfnamefont {N.}~\bibnamefont {Kiesel}}, \bibinfo {author}
  {\bibfnamefont {M.}~\bibnamefont {Aspelmeyer}},\ and\ \bibinfo {author}
  {\bibfnamefont {J.~I.}\ \bibnamefont {Cirac}},\ }\bibfield  {title} {\bibinfo
  {title} {Large quantum superpositions and interference of massive
  nanometer-sized objects},\ }\href
  {https://doi.org/10.1103/PhysRevLett.107.020405} {\bibfield  {journal}
  {\bibinfo  {journal} {Phys. Rev. Lett.}\ }\textbf {\bibinfo {volume} {107}},\
  \bibinfo {pages} {020405} (\bibinfo {year} {2011})}\BibitemShut {NoStop}%
\bibitem [{\citenamefont {Romero-Isart}(2011)}]{oriol_collapsemodels_2011}%
  \BibitemOpen
  \bibfield  {author} {\bibinfo {author} {\bibfnamefont {O.}~\bibnamefont
  {Romero-Isart}},\ }\bibfield  {title} {\bibinfo {title} {Quantum
  superposition of massive objects and collapse models},\ }\href
  {https://doi.org/10.1103/PhysRevA.84.052121} {\bibfield  {journal} {\bibinfo
  {journal} {Phys. Rev. A}\ }\textbf {\bibinfo {volume} {84}},\ \bibinfo
  {pages} {052121} (\bibinfo {year} {2011})}\BibitemShut {NoStop}%
\bibitem [{\citenamefont {Fonseca}\ \emph {et~al.}(2016)\citenamefont
  {Fonseca}, \citenamefont {Aranas}, \citenamefont {Millen}, \citenamefont
  {Monteiro},\ and\ \citenamefont {Barker}}]{Fonseca_2016}%
  \BibitemOpen
  \bibfield  {author} {\bibinfo {author} {\bibfnamefont {P.~Z.~G.}\
  \bibnamefont {Fonseca}}, \bibinfo {author} {\bibfnamefont {E.~B.}\
  \bibnamefont {Aranas}}, \bibinfo {author} {\bibfnamefont {J.}~\bibnamefont
  {Millen}}, \bibinfo {author} {\bibfnamefont {T.~S.}\ \bibnamefont
  {Monteiro}},\ and\ \bibinfo {author} {\bibfnamefont {P.~F.}\ \bibnamefont
  {Barker}},\ }\bibfield  {title} {\bibinfo {title} {Nonlinear dynamics and
  strong cavity cooling of levitated nanoparticles},\ }\href
  {https://doi.org/10.1103/PhysRevLett.117.173602} {\bibfield  {journal}
  {\bibinfo  {journal} {Phys. Rev. Lett.}\ }\textbf {\bibinfo {volume} {117}},\
  \bibinfo {pages} {173602} (\bibinfo {year} {2016})}\BibitemShut {NoStop}%
\bibitem [{\citenamefont {Goldwater}\ \emph {et~al.}(2019)\citenamefont
  {Goldwater}, \citenamefont {Stickler}, \citenamefont {Martinetz},
  \citenamefont {Northup}, \citenamefont {Hornberger},\ and\ \citenamefont
  {Millen}}]{Goldwater_2019}%
  \BibitemOpen
  \bibfield  {author} {\bibinfo {author} {\bibfnamefont {D.}~\bibnamefont
  {Goldwater}}, \bibinfo {author} {\bibfnamefont {B.~A.}\ \bibnamefont
  {Stickler}}, \bibinfo {author} {\bibfnamefont {L.}~\bibnamefont {Martinetz}},
  \bibinfo {author} {\bibfnamefont {T.~E.}\ \bibnamefont {Northup}}, \bibinfo
  {author} {\bibfnamefont {K.}~\bibnamefont {Hornberger}},\ and\ \bibinfo
  {author} {\bibfnamefont {J.}~\bibnamefont {Millen}},\ }\bibfield  {title}
  {\bibinfo {title} {Levitated electromechanics: all-electrical cooling of
  charged nano- and micro-particles},\ }\href
  {https://doi.org/10.1088/2058-9565/aaf5f3} {\bibfield  {journal} {\bibinfo
  {journal} {Quantum Sci. Technol.}\ }\textbf {\bibinfo {volume} {4}},\
  \bibinfo {pages} {024003} (\bibinfo {year} {2019})}\BibitemShut {NoStop}%
\bibitem [{\citenamefont {Martinetz}\ \emph {et~al.}(2020)\citenamefont
  {Martinetz}, \citenamefont {Hornberger}, \citenamefont {Millen},
  \citenamefont {Kim},\ and\ \citenamefont {Stickler}}]{Martinetz2020}%
  \BibitemOpen
  \bibfield  {author} {\bibinfo {author} {\bibfnamefont {L.}~\bibnamefont
  {Martinetz}}, \bibinfo {author} {\bibfnamefont {K.}~\bibnamefont
  {Hornberger}}, \bibinfo {author} {\bibfnamefont {J.}~\bibnamefont {Millen}},
  \bibinfo {author} {\bibfnamefont {M.~S.}\ \bibnamefont {Kim}},\ and\ \bibinfo
  {author} {\bibfnamefont {B.~A.}\ \bibnamefont {Stickler}},\ }\bibfield
  {title} {\bibinfo {title} {Quantum electromechanics with levitated
  nanoparticles},\ }\href {https://doi.org/10.1038/s41534-020-00333-7}
  {\bibfield  {journal} {\bibinfo  {journal} {Npj Quantum Inf.}\ }\textbf
  {\bibinfo {volume} {6}},\ \bibinfo {pages} {1} (\bibinfo {year}
  {2020})}\BibitemShut {NoStop}%
\bibitem [{\citenamefont {Dania}\ \emph {et~al.}(2021)\citenamefont {Dania},
  \citenamefont {Bykov}, \citenamefont {Knoll}, \citenamefont {Mestres},\ and\
  \citenamefont {Northup}}]{Lorenzo2021}%
  \BibitemOpen
  \bibfield  {author} {\bibinfo {author} {\bibfnamefont {L.}~\bibnamefont
  {Dania}}, \bibinfo {author} {\bibfnamefont {D.~S.}\ \bibnamefont {Bykov}},
  \bibinfo {author} {\bibfnamefont {M.}~\bibnamefont {Knoll}}, \bibinfo
  {author} {\bibfnamefont {P.}~\bibnamefont {Mestres}},\ and\ \bibinfo {author}
  {\bibfnamefont {T.~E.}\ \bibnamefont {Northup}},\ }\bibfield  {title}
  {\bibinfo {title} {Optical and electrical feedback cooling of a silica
  nanoparticle levitated in a paul trap},\ }\href
  {https://doi.org/10.1103/PhysRevResearch.3.013018} {\bibfield  {journal}
  {\bibinfo  {journal} {Phys. Rev. Research}\ }\textbf {\bibinfo {volume}
  {3}},\ \bibinfo {pages} {013018} (\bibinfo {year} {2021})}\BibitemShut
  {NoStop}%
\bibitem [{\citenamefont {Moon}\ and\ \citenamefont
  {Chang}(1994)}]{moon_superconducting_1994}%
  \BibitemOpen
  \bibfield  {author} {\bibinfo {author} {\bibfnamefont {F.}~\bibnamefont
  {Moon}}\ and\ \bibinfo {author} {\bibfnamefont {P.}~\bibnamefont {Chang}},\
  }\href {https://books.google.at/books?id=ARdRAAAAMAAJ} {\emph {\bibinfo
  {title} {Superconducting {Levitation}: {Applications} to {Bearings} and
  {Magnetic} {Transportation}}}},\ A {Wiley} interscience publication\
  (\bibinfo  {publisher} {Wiley},\ \bibinfo {year} {1994})\BibitemShut
  {NoStop}%
\bibitem [{\citenamefont {Takahashi}\ \emph {et~al.}(2017)\citenamefont
  {Takahashi}, \citenamefont {Suzuki}, \citenamefont {Yoneyama}, \citenamefont
  {Tokawa}, \citenamefont {Suzuki}, \citenamefont {Matsushima}, \citenamefont
  {Kumakura}, \citenamefont {Ashida},\ and\ \citenamefont
  {Moriwaki}}]{Takahashi2017}%
  \BibitemOpen
  \bibfield  {author} {\bibinfo {author} {\bibfnamefont {Y.}~\bibnamefont
  {Takahashi}}, \bibinfo {author} {\bibfnamefont {J.}~\bibnamefont {Suzuki}},
  \bibinfo {author} {\bibfnamefont {N.}~\bibnamefont {Yoneyama}}, \bibinfo
  {author} {\bibfnamefont {Y.}~\bibnamefont {Tokawa}}, \bibinfo {author}
  {\bibfnamefont {N.}~\bibnamefont {Suzuki}}, \bibinfo {author} {\bibfnamefont
  {F.}~\bibnamefont {Matsushima}}, \bibinfo {author} {\bibfnamefont
  {M.}~\bibnamefont {Kumakura}}, \bibinfo {author} {\bibfnamefont
  {M.}~\bibnamefont {Ashida}},\ and\ \bibinfo {author} {\bibfnamefont
  {Y.}~\bibnamefont {Moriwaki}},\ }\bibfield  {title} {\bibinfo {title}
  {Magnetic trapping of superconducting submicron particles produced by laser
  ablation in superfluid helium},\ }\href
  {https://doi.org/10.7567/APEX.10.022701} {\bibfield  {journal} {\bibinfo
  {journal} {Appl. Phys. Express}\ }\textbf {\bibinfo {volume} {10}},\ \bibinfo
  {pages} {22701} (\bibinfo {year} {2017})}\BibitemShut {NoStop}%
\bibitem [{\citenamefont {Vinante}\ \emph {et~al.}(2020)\citenamefont
  {Vinante}, \citenamefont {Falferi}, \citenamefont {Gasbarri}, \citenamefont
  {Setter}, \citenamefont {Timberlake},\ and\ \citenamefont
  {Ulbricht}}]{Vinante_2020}%
  \BibitemOpen
  \bibfield  {author} {\bibinfo {author} {\bibfnamefont {A.}~\bibnamefont
  {Vinante}}, \bibinfo {author} {\bibfnamefont {P.}~\bibnamefont {Falferi}},
  \bibinfo {author} {\bibfnamefont {G.}~\bibnamefont {Gasbarri}}, \bibinfo
  {author} {\bibfnamefont {A.}~\bibnamefont {Setter}}, \bibinfo {author}
  {\bibfnamefont {C.}~\bibnamefont {Timberlake}},\ and\ \bibinfo {author}
  {\bibfnamefont {H.}~\bibnamefont {Ulbricht}},\ }\bibfield  {title} {\bibinfo
  {title} {Ultralow mechanical damping with meissner-levitated ferromagnetic
  microparticles},\ }\href {https://doi.org/10.1103/PhysRevApplied.13.064027}
  {\bibfield  {journal} {\bibinfo  {journal} {Phys. Rev. Appl.}\ }\textbf
  {\bibinfo {volume} {13}},\ \bibinfo {pages} {064027} (\bibinfo {year}
  {2020})}\BibitemShut {NoStop}%
\bibitem [{\citenamefont {Zheng}\ \emph
  {et~al.}(2020{\natexlab{a}})\citenamefont {Zheng}, \citenamefont {Leng},
  \citenamefont {Kong}, \citenamefont {Li}, \citenamefont {Wang}, \citenamefont
  {Luo}, \citenamefont {Zhao}, \citenamefont {Duan}, \citenamefont {Huang},
  \citenamefont {Du}, \citenamefont {Carlesso},\ and\ \citenamefont
  {Bassi}}]{zheng_room_2020}%
  \BibitemOpen
  \bibfield  {author} {\bibinfo {author} {\bibfnamefont {D.}~\bibnamefont
  {Zheng}}, \bibinfo {author} {\bibfnamefont {Y.}~\bibnamefont {Leng}},
  \bibinfo {author} {\bibfnamefont {X.}~\bibnamefont {Kong}}, \bibinfo {author}
  {\bibfnamefont {R.}~\bibnamefont {Li}}, \bibinfo {author} {\bibfnamefont
  {Z.}~\bibnamefont {Wang}}, \bibinfo {author} {\bibfnamefont {X.}~\bibnamefont
  {Luo}}, \bibinfo {author} {\bibfnamefont {J.}~\bibnamefont {Zhao}}, \bibinfo
  {author} {\bibfnamefont {C.-K.}\ \bibnamefont {Duan}}, \bibinfo {author}
  {\bibfnamefont {P.}~\bibnamefont {Huang}}, \bibinfo {author} {\bibfnamefont
  {J.}~\bibnamefont {Du}}, \bibinfo {author} {\bibfnamefont {M.}~\bibnamefont
  {Carlesso}},\ and\ \bibinfo {author} {\bibfnamefont {A.}~\bibnamefont
  {Bassi}},\ }\bibfield  {title} {\bibinfo {title} {Room temperature test of
  the continuous spontaneous localization model using a levitated
  micro-oscillator},\ }\href {https://doi.org/10.1103/PhysRevResearch.2.013057}
  {\bibfield  {journal} {\bibinfo  {journal} {Phys. Rev. Research}\ }\textbf
  {\bibinfo {volume} {2}},\ \bibinfo {pages} {013057} (\bibinfo {year}
  {2020}{\natexlab{a}})}\BibitemShut {NoStop}%
\bibitem [{\citenamefont {Leng}\ \emph {et~al.}(2021)\citenamefont {Leng},
  \citenamefont {Li}, \citenamefont {Kong}, \citenamefont {Xie}, \citenamefont
  {Zheng}, \citenamefont {Yin}, \citenamefont {Xiong}, \citenamefont {Wu},
  \citenamefont {Duan}, \citenamefont {Du},\ and\ \citenamefont
  {et~al.}}]{Leng_2021}%
  \BibitemOpen
  \bibfield  {author} {\bibinfo {author} {\bibfnamefont {Y.}~\bibnamefont
  {Leng}}, \bibinfo {author} {\bibfnamefont {R.}~\bibnamefont {Li}}, \bibinfo
  {author} {\bibfnamefont {X.}~\bibnamefont {Kong}}, \bibinfo {author}
  {\bibfnamefont {H.}~\bibnamefont {Xie}}, \bibinfo {author} {\bibfnamefont
  {D.}~\bibnamefont {Zheng}}, \bibinfo {author} {\bibfnamefont
  {P.}~\bibnamefont {Yin}}, \bibinfo {author} {\bibfnamefont {F.}~\bibnamefont
  {Xiong}}, \bibinfo {author} {\bibfnamefont {T.}~\bibnamefont {Wu}}, \bibinfo
  {author} {\bibfnamefont {C.-K.}\ \bibnamefont {Duan}}, \bibinfo {author}
  {\bibfnamefont {Y.}~\bibnamefont {Du}},\ and\ \bibinfo {author} {\bibnamefont
  {et~al.}},\ }\bibfield  {title} {\bibinfo {title} {Mechanical dissipation
  below 1 muhz with a cryogenic diamagnetic levitated micro-oscillator},\
  }\href {https://doi.org/10.1103/PhysRevApplied.15.024061} {\bibfield
  {journal} {\bibinfo  {journal} {Phys. Rev. Appl.}\ }\textbf {\bibinfo
  {volume} {15}},\ \bibinfo {pages} {024061} (\bibinfo {year}
  {2021})}\BibitemShut {NoStop}%
\bibitem [{\citenamefont {Brown}\ \emph {et~al.}(2021)\citenamefont {Brown},
  \citenamefont {Wang}, \citenamefont {Namazi}, \citenamefont {Harris},
  \citenamefont {Uysal},\ and\ \citenamefont {Harris}}]{Brown2021}%
  \BibitemOpen
  \bibfield  {author} {\bibinfo {author} {\bibfnamefont {C.~D.}\ \bibnamefont
  {Brown}}, \bibinfo {author} {\bibfnamefont {Y.}~\bibnamefont {Wang}},
  \bibinfo {author} {\bibfnamefont {M.}~\bibnamefont {Namazi}}, \bibinfo
  {author} {\bibfnamefont {G.~I.}\ \bibnamefont {Harris}}, \bibinfo {author}
  {\bibfnamefont {M.~T.}\ \bibnamefont {Uysal}},\ and\ \bibinfo {author}
  {\bibfnamefont {J.~G.~E.}\ \bibnamefont {Harris}},\ }\href
  {https://doi.org/10.48550/arXiv.2109.05618} {\bibinfo {title} {Superfluid
  helium drops levitated in high vacuum}} (\bibinfo {year} {2021}),\ \Eprint
  {https://arxiv.org/abs/arXiv:2109.05618} {arXiv:2109.05618} \BibitemShut
  {NoStop}%
\bibitem [{\citenamefont {Array{\'{a}}s}\ \emph {et~al.}(2021)\citenamefont
  {Array{\'{a}}s}, \citenamefont {Trueba}, \citenamefont {Uriarte},\ and\
  \citenamefont {Zmeev}}]{dima_carlos_2021}%
  \BibitemOpen
  \bibfield  {author} {\bibinfo {author} {\bibfnamefont {M.}~\bibnamefont
  {Array{\'{a}}s}}, \bibinfo {author} {\bibfnamefont {J.~L.}\ \bibnamefont
  {Trueba}}, \bibinfo {author} {\bibfnamefont {C.}~\bibnamefont {Uriarte}},\
  and\ \bibinfo {author} {\bibfnamefont {D.~E.}\ \bibnamefont {Zmeev}},\
  }\bibfield  {title} {\bibinfo {title} {Design of a system for controlling a
  levitating sphere in superfluid 3he at extremely low temperatures},\ }\href
  {https://doi.org/10.1038/s41598-021-99316-7} {\bibfield  {journal} {\bibinfo
  {journal} {Sci. Rep. 2021 11:1}\ }\textbf {\bibinfo {volume} {11}},\ \bibinfo
  {pages} {1} (\bibinfo {year} {2021})}\BibitemShut {NoStop}%
\bibitem [{\citenamefont {Hofer}\ \emph {et~al.}(2022)\citenamefont {Hofer},
  \citenamefont {Higgins}, \citenamefont {Huebl}, \citenamefont {Kieler},
  \citenamefont {Kleiner}, \citenamefont {Koelle}, \citenamefont {Schmidt},
  \citenamefont {Slater}, \citenamefont {Trupke}, \citenamefont {Uhl},
  \citenamefont {Weimann}, \citenamefont {Wieczorek}, \citenamefont
  {Wulschner},\ and\ \citenamefont
  {Aspelmeyer}}]{hoferHighQMagneticLevitation2022}%
  \BibitemOpen
  \bibfield  {author} {\bibinfo {author} {\bibfnamefont {J.}~\bibnamefont
  {Hofer}}, \bibinfo {author} {\bibfnamefont {G.}~\bibnamefont {Higgins}},
  \bibinfo {author} {\bibfnamefont {H.}~\bibnamefont {Huebl}}, \bibinfo
  {author} {\bibfnamefont {O.~F.}\ \bibnamefont {Kieler}}, \bibinfo {author}
  {\bibfnamefont {R.}~\bibnamefont {Kleiner}}, \bibinfo {author} {\bibfnamefont
  {D.}~\bibnamefont {Koelle}}, \bibinfo {author} {\bibfnamefont
  {P.}~\bibnamefont {Schmidt}}, \bibinfo {author} {\bibfnamefont {J.~A.}\
  \bibnamefont {Slater}}, \bibinfo {author} {\bibfnamefont {M.}~\bibnamefont
  {Trupke}}, \bibinfo {author} {\bibfnamefont {K.}~\bibnamefont {Uhl}},
  \bibinfo {author} {\bibfnamefont {T.}~\bibnamefont {Weimann}}, \bibinfo
  {author} {\bibfnamefont {W.}~\bibnamefont {Wieczorek}}, \bibinfo {author}
  {\bibfnamefont {F.}~\bibnamefont {Wulschner}},\ and\ \bibinfo {author}
  {\bibfnamefont {M.}~\bibnamefont {Aspelmeyer}},\ }\href
  {https://doi.org/10.48550/arXiv.2211.06289} {\bibinfo {title} {High-{{Q}}
  magnetic levitation and control of superconducting microspheres at
  millikelvin temperatures}} (\bibinfo {year} {2022}),\ \Eprint
  {https://arxiv.org/abs/arXiv:2211.06289} {arXiv:2211.06289} \BibitemShut
  {NoStop}%
\bibitem [{\citenamefont {Latorre}\ \emph {et~al.}(2022)\citenamefont
  {Latorre}, \citenamefont {Paradkar}, \citenamefont {Hambraeus}, \citenamefont
  {Higgins},\ and\ \citenamefont {Wieczorek}}]{marti_2022}%
  \BibitemOpen
  \bibfield  {author} {\bibinfo {author} {\bibfnamefont {M.~G.}\ \bibnamefont
  {Latorre}}, \bibinfo {author} {\bibfnamefont {A.}~\bibnamefont {Paradkar}},
  \bibinfo {author} {\bibfnamefont {D.}~\bibnamefont {Hambraeus}}, \bibinfo
  {author} {\bibfnamefont {G.}~\bibnamefont {Higgins}},\ and\ \bibinfo {author}
  {\bibfnamefont {W.}~\bibnamefont {Wieczorek}},\ }\bibfield  {title} {\bibinfo
  {title} {A chip-based superconducting magnetic trap for levitating
  superconducting microparticles},\ }\href
  {https://doi.org/10.1109/TASC.2022.3147730} {\bibfield  {journal} {\bibinfo
  {journal} {IEEE Trans. Appl. Supercond}\ }\textbf {\bibinfo {volume} {32}},\
  \bibinfo {pages} {1} (\bibinfo {year} {2022})}\BibitemShut {NoStop}%
\bibitem [{\citenamefont {Latorre}\ \emph {et~al.}(2020)\citenamefont
  {Latorre}, \citenamefont {Hofer}, \citenamefont {Rudolph},\ and\
  \citenamefont {Wieczorek}}]{marti_2020}%
  \BibitemOpen
  \bibfield  {author} {\bibinfo {author} {\bibfnamefont {M.~G.}\ \bibnamefont
  {Latorre}}, \bibinfo {author} {\bibfnamefont {J.}~\bibnamefont {Hofer}},
  \bibinfo {author} {\bibfnamefont {M.}~\bibnamefont {Rudolph}},\ and\ \bibinfo
  {author} {\bibfnamefont {W.}~\bibnamefont {Wieczorek}},\ }\bibfield  {title}
  {\bibinfo {title} {Chip-based superconducting traps for levitation of
  micrometer-sized particles in the meissner state},\ }\href
  {https://doi.org/10.1088/1361-6668/aba6e1} {\bibfield  {journal} {\bibinfo
  {journal} {Supercond. Sci. Technol.}\ }\textbf {\bibinfo {volume} {33}},\
  \bibinfo {pages} {105002} (\bibinfo {year} {2020})}\BibitemShut {NoStop}%
\bibitem [{\citenamefont {Navau}\ \emph {et~al.}(2021)\citenamefont {Navau},
  \citenamefont {Minniberger}, \citenamefont {Trupke},\ and\ \citenamefont
  {Sanchez}}]{Navau_2021}%
  \BibitemOpen
  \bibfield  {author} {\bibinfo {author} {\bibfnamefont {C.}~\bibnamefont
  {Navau}}, \bibinfo {author} {\bibfnamefont {S.}~\bibnamefont {Minniberger}},
  \bibinfo {author} {\bibfnamefont {M.}~\bibnamefont {Trupke}},\ and\ \bibinfo
  {author} {\bibfnamefont {A.}~\bibnamefont {Sanchez}},\ }\bibfield  {title}
  {\bibinfo {title} {Levitation of superconducting microrings for quantum
  magnetomechanics},\ }\href {https://doi.org/10.1103/PhysRevB.103.174436}
  {\bibfield  {journal} {\bibinfo  {journal} {Phys. Rev. B}\ }\textbf {\bibinfo
  {volume} {103}},\ \bibinfo {pages} {174436} (\bibinfo {year}
  {2021})}\BibitemShut {NoStop}%
\bibitem [{\citenamefont {Romero-Isart}\ \emph {et~al.}(2012)\citenamefont
  {Romero-Isart}, \citenamefont {Clemente}, \citenamefont {Navau},
  \citenamefont {Sanchez},\ and\ \citenamefont {Cirac}}]{oriol}%
  \BibitemOpen
  \bibfield  {author} {\bibinfo {author} {\bibfnamefont {O.}~\bibnamefont
  {Romero-Isart}}, \bibinfo {author} {\bibfnamefont {L.}~\bibnamefont
  {Clemente}}, \bibinfo {author} {\bibfnamefont {C.}~\bibnamefont {Navau}},
  \bibinfo {author} {\bibfnamefont {A.}~\bibnamefont {Sanchez}},\ and\ \bibinfo
  {author} {\bibfnamefont {J.~I.}\ \bibnamefont {Cirac}},\ }\bibfield  {title}
  {\bibinfo {title} {Quantum magnetomechanics with levitating superconducting
  microspheres},\ }\href {https://doi.org/10.1103/PhysRevLett.109.147205}
  {\bibfield  {journal} {\bibinfo  {journal} {Phys. Rev. Lett.}\ }\textbf
  {\bibinfo {volume} {109}},\ \bibinfo {pages} {147205} (\bibinfo {year}
  {2012})}\BibitemShut {NoStop}%
\bibitem [{\citenamefont {Cirio}\ \emph {et~al.}(2012)\citenamefont {Cirio},
  \citenamefont {Brennen},\ and\ \citenamefont {Twamley}}]{cirio_quantum_2012}%
  \BibitemOpen
  \bibfield  {author} {\bibinfo {author} {\bibfnamefont {M.}~\bibnamefont
  {Cirio}}, \bibinfo {author} {\bibfnamefont {G.~K.}\ \bibnamefont {Brennen}},\
  and\ \bibinfo {author} {\bibfnamefont {J.}~\bibnamefont {Twamley}},\
  }\bibfield  {title} {\bibinfo {title} {Quantum {Magnetomechanics}:
  {Ultrahigh}- {Q} -{Levitated} {Mechanical} {Oscillators}},\ }\href
  {https://doi.org/10.1103/PhysRevLett.109.147206} {\bibfield  {journal}
  {\bibinfo  {journal} {Phys. Rev. Lett.}\ }\textbf {\bibinfo {volume} {109}},\
  \bibinfo {pages} {147206} (\bibinfo {year} {2012})}\BibitemShut {NoStop}%
\bibitem [{\citenamefont {Pino}\ \emph {et~al.}(2018)\citenamefont {Pino},
  \citenamefont {Prat-Camps}, \citenamefont {Sinha}, \citenamefont
  {Venkatesh},\ and\ \citenamefont {Romero-Isart}}]{Pino_2018}%
  \BibitemOpen
  \bibfield  {author} {\bibinfo {author} {\bibfnamefont {H.}~\bibnamefont
  {Pino}}, \bibinfo {author} {\bibfnamefont {J.}~\bibnamefont {Prat-Camps}},
  \bibinfo {author} {\bibfnamefont {K.}~\bibnamefont {Sinha}}, \bibinfo
  {author} {\bibfnamefont {B.~P.}\ \bibnamefont {Venkatesh}},\ and\ \bibinfo
  {author} {\bibfnamefont {O.}~\bibnamefont {Romero-Isart}},\ }\bibfield
  {title} {\bibinfo {title} {On-chip quantum interference of a superconducting
  microsphere},\ }\href {https://doi.org/10.1088/2058-9565/aa9d15} {\bibfield
  {journal} {\bibinfo  {journal} {Quantum Sci. Technol.}\ }\textbf {\bibinfo
  {volume} {3}},\ \bibinfo {pages} {025001} (\bibinfo {year}
  {2018})}\BibitemShut {NoStop}%
\bibitem [{\citenamefont {Jackson~Kimball}\ \emph {et~al.}(2016)\citenamefont
  {Jackson~Kimball}, \citenamefont {Sushkov},\ and\ \citenamefont
  {Budker}}]{jacksonkimballPrecessingFerromagneticNeedle2016}%
  \BibitemOpen
  \bibfield  {author} {\bibinfo {author} {\bibfnamefont {D.~F.}\ \bibnamefont
  {Jackson~Kimball}}, \bibinfo {author} {\bibfnamefont {A.~O.}\ \bibnamefont
  {Sushkov}},\ and\ \bibinfo {author} {\bibfnamefont {D.}~\bibnamefont
  {Budker}},\ }\bibfield  {title} {\bibinfo {title} {Precessing {{Ferromagnetic
  Needle Magnetometer}}},\ }\href
  {https://doi.org/10.1103/PhysRevLett.116.190801} {\bibfield  {journal}
  {\bibinfo  {journal} {Phys. Rev. Lett.}\ }\textbf {\bibinfo {volume} {116}},\
  \bibinfo {pages} {190801} (\bibinfo {year} {2016})}\BibitemShut {NoStop}%
\bibitem [{\citenamefont {Niemetz}\ \emph {et~al.}(2000)\citenamefont
  {Niemetz}, \citenamefont {Schoepe}, \citenamefont {Simola},\ and\
  \citenamefont {Tuoriniemi}}]{niemetz_oscillating_2000}%
  \BibitemOpen
  \bibfield  {author} {\bibinfo {author} {\bibfnamefont {M.}~\bibnamefont
  {Niemetz}}, \bibinfo {author} {\bibfnamefont {W.}~\bibnamefont {Schoepe}},
  \bibinfo {author} {\bibfnamefont {J.~T.}\ \bibnamefont {Simola}},\ and\
  \bibinfo {author} {\bibfnamefont {J.~T.}\ \bibnamefont {Tuoriniemi}},\
  }\bibfield  {title} {\bibinfo {title} {The oscillating magnetic microsphere:
  a tool for investigating vorticity in superconductors and superfluids},\
  }\href {https://doi.org/http://dx.doi.org/10.1016/S0921-4526(99)01864-5}
  {\bibfield  {journal} {\bibinfo  {journal} {Phys. B: Condens. Matter}\
  }\textbf {\bibinfo {volume} {280}},\ \bibinfo {pages} {559 } (\bibinfo {year}
  {2000})}\BibitemShut {NoStop}%
\bibitem [{\citenamefont {Wang}\ \emph {et~al.}(2019)\citenamefont {Wang},
  \citenamefont {Lourette}, \citenamefont {O'Kelley}, \citenamefont {Kayci},
  \citenamefont {Band}, \citenamefont {Kimball}, \citenamefont {Sushkov},\ and\
  \citenamefont {Budker}}]{Wang2019}%
  \BibitemOpen
  \bibfield  {author} {\bibinfo {author} {\bibfnamefont {T.}~\bibnamefont
  {Wang}}, \bibinfo {author} {\bibfnamefont {S.}~\bibnamefont {Lourette}},
  \bibinfo {author} {\bibfnamefont {S.~R.}\ \bibnamefont {O'Kelley}}, \bibinfo
  {author} {\bibfnamefont {M.}~\bibnamefont {Kayci}}, \bibinfo {author}
  {\bibfnamefont {Y.~B.}\ \bibnamefont {Band}}, \bibinfo {author}
  {\bibfnamefont {D.~F.~J.}\ \bibnamefont {Kimball}}, \bibinfo {author}
  {\bibfnamefont {A.~O.}\ \bibnamefont {Sushkov}},\ and\ \bibinfo {author}
  {\bibfnamefont {D.}~\bibnamefont {Budker}},\ }\bibfield  {title} {\bibinfo
  {title} {Dynamics of a ferromagnetic particle levitated over a
  superconductor},\ }\href {https://doi.org/10.1103/PhysRevApplied.11.044041}
  {\bibfield  {journal} {\bibinfo  {journal} {Phys. Rev. Applied}\ }\textbf
  {\bibinfo {volume} {11}},\ \bibinfo {pages} {044041} (\bibinfo {year}
  {2019})}\BibitemShut {NoStop}%
\bibitem [{\citenamefont {Gieseler}\ \emph {et~al.}(2020)\citenamefont
  {Gieseler}, \citenamefont {Kabcenell}, \citenamefont {Rosenfeld},
  \citenamefont {Schaefer}, \citenamefont {Safira}, \citenamefont {Schuetz},
  \citenamefont {Gonzalez-Ballestero}, \citenamefont {Rusconi}, \citenamefont
  {Romero-Isart},\ and\ \citenamefont {Lukin}}]{Gieseler_2020}%
  \BibitemOpen
  \bibfield  {author} {\bibinfo {author} {\bibfnamefont {J.}~\bibnamefont
  {Gieseler}}, \bibinfo {author} {\bibfnamefont {A.}~\bibnamefont {Kabcenell}},
  \bibinfo {author} {\bibfnamefont {E.}~\bibnamefont {Rosenfeld}}, \bibinfo
  {author} {\bibfnamefont {J.~D.}\ \bibnamefont {Schaefer}}, \bibinfo {author}
  {\bibfnamefont {A.}~\bibnamefont {Safira}}, \bibinfo {author} {\bibfnamefont
  {M.~J.~A.}\ \bibnamefont {Schuetz}}, \bibinfo {author} {\bibfnamefont
  {C.}~\bibnamefont {Gonzalez-Ballestero}}, \bibinfo {author} {\bibfnamefont
  {C.~C.}\ \bibnamefont {Rusconi}}, \bibinfo {author} {\bibfnamefont
  {O.}~\bibnamefont {Romero-Isart}},\ and\ \bibinfo {author} {\bibfnamefont
  {M.~D.}\ \bibnamefont {Lukin}},\ }\bibfield  {title} {\bibinfo {title}
  {Single-spin magnetomechanics with levitated micromagnets},\ }\href
  {https://doi.org/10.1103/PhysRevLett.124.163604} {\bibfield  {journal}
  {\bibinfo  {journal} {Phys. Rev. Lett.}\ }\textbf {\bibinfo {volume} {124}},\
  \bibinfo {pages} {163604} (\bibinfo {year} {2020})}\BibitemShut {NoStop}%
\bibitem [{\citenamefont {Slezak}\ \emph {et~al.}(2018)\citenamefont {Slezak},
  \citenamefont {Lewandowski}, \citenamefont {Hsu},\ and\ \citenamefont
  {D’Urso}}]{Slezak_2018}%
  \BibitemOpen
  \bibfield  {author} {\bibinfo {author} {\bibfnamefont {B.~R.}\ \bibnamefont
  {Slezak}}, \bibinfo {author} {\bibfnamefont {C.~W.}\ \bibnamefont
  {Lewandowski}}, \bibinfo {author} {\bibfnamefont {J.-F.}\ \bibnamefont
  {Hsu}},\ and\ \bibinfo {author} {\bibfnamefont {B.}~\bibnamefont
  {D’Urso}},\ }\bibfield  {title} {\bibinfo {title} {Cooling the motion of a
  silica microsphere in a magneto-gravitational trap in ultra-high vacuum},\
  }\href {https://doi.org/10.1088/1367-2630/aacac1} {\bibfield  {journal}
  {\bibinfo  {journal} {New J. Phys.}\ }\textbf {\bibinfo {volume} {20}},\
  \bibinfo {pages} {063028} (\bibinfo {year} {2018})}\BibitemShut {NoStop}%
\bibitem [{\citenamefont {van Waarde}(2016)}]{Waarde_2016}%
  \BibitemOpen
  \bibfield  {author} {\bibinfo {author} {\bibfnamefont {B.}~\bibnamefont {van
  Waarde}},\ }\emph {\bibinfo {title} {The lead zeppelin - a force sensor
  without a handle}},\ \href
  {https://scholarlypublications.universiteitleiden.nl/handle/1887/43816}
  {Ph.D. thesis},\ \bibinfo  {school} {Universiteit Leiden, The Netherlands}
  (\bibinfo {year} {2016})\BibitemShut {NoStop}%
\bibitem [{\citenamefont {{van Waarde}}\ \emph {et~al.}(2016)\citenamefont
  {{van Waarde}}, \citenamefont {Benningshof},\ and\ \citenamefont
  {Oosterkamp}}]{vanwaardeMagneticPersistentCurrent2016}%
  \BibitemOpen
  \bibfield  {author} {\bibinfo {author} {\bibfnamefont {B.}~\bibnamefont {{van
  Waarde}}}, \bibinfo {author} {\bibfnamefont {O.}~\bibnamefont
  {Benningshof}},\ and\ \bibinfo {author} {\bibfnamefont {T.}~\bibnamefont
  {Oosterkamp}},\ }\bibfield  {title} {\bibinfo {title} {A magnetic persistent
  current switch at {{milliKelvin}} temperatures},\ }\href
  {https://doi.org/10.1016/j.cryogenics.2016.06.014} {\bibfield  {journal}
  {\bibinfo  {journal} {Cryogenics}\ }\textbf {\bibinfo {volume} {78}},\
  \bibinfo {pages} {74} (\bibinfo {year} {2016})}\BibitemShut {NoStop}%
\bibitem [{\citenamefont {Hsu}\ \emph {et~al.}(2016)\citenamefont {Hsu},
  \citenamefont {Ji}, \citenamefont {Lewandowski},\ and\ \citenamefont
  {D'Urso}}]{Hsu2016}%
  \BibitemOpen
  \bibfield  {author} {\bibinfo {author} {\bibfnamefont {J.~F.}\ \bibnamefont
  {Hsu}}, \bibinfo {author} {\bibfnamefont {P.}~\bibnamefont {Ji}}, \bibinfo
  {author} {\bibfnamefont {C.~W.}\ \bibnamefont {Lewandowski}},\ and\ \bibinfo
  {author} {\bibfnamefont {B.}~\bibnamefont {D'Urso}},\ }\bibfield  {title}
  {\bibinfo {title} {Cooling the motion of diamond nanocrystals in a
  magneto-gravitational trap in high vacuum},\ }\href
  {https://doi.org/10.1038/srep30125} {\bibfield  {journal} {\bibinfo
  {journal} {Sci. Rep.}\ }\textbf {\bibinfo {volume} {6}},\ \bibinfo {pages}
  {1} (\bibinfo {year} {2016})}\BibitemShut {NoStop}%
\bibitem [{\citenamefont {Weinstein}\ and\ \citenamefont
  {Libbrecht}(1995)}]{Weinstein1995}%
  \BibitemOpen
  \bibfield  {author} {\bibinfo {author} {\bibfnamefont {J.~D.}\ \bibnamefont
  {Weinstein}}\ and\ \bibinfo {author} {\bibfnamefont {K.~G.}\ \bibnamefont
  {Libbrecht}},\ }\bibfield  {title} {\bibinfo {title} {Microscopic magnetic
  traps for neutral atoms},\ }\href {https://doi.org/10.1103/PhysRevA.52.4004}
  {\bibfield  {journal} {\bibinfo  {journal} {Phys. Rev. A}\ }\textbf {\bibinfo
  {volume} {52}},\ \bibinfo {pages} {4004} (\bibinfo {year}
  {1995})}\BibitemShut {NoStop}%
\bibitem [{\citenamefont {Reichel}\ \emph {et~al.}(1999)\citenamefont
  {Reichel}, \citenamefont {H\"ansel},\ and\ \citenamefont
  {H\"ansch}}]{Reichel1999}%
  \BibitemOpen
  \bibfield  {author} {\bibinfo {author} {\bibfnamefont {J.}~\bibnamefont
  {Reichel}}, \bibinfo {author} {\bibfnamefont {W.}~\bibnamefont {H\"ansel}},\
  and\ \bibinfo {author} {\bibfnamefont {T.~W.}\ \bibnamefont {H\"ansch}},\
  }\bibfield  {title} {\bibinfo {title} {Atomic micromanipulation with magnetic
  surface traps},\ }\href {https://doi.org/10.1103/PhysRevLett.83.3398}
  {\bibfield  {journal} {\bibinfo  {journal} {Phys. Rev. Lett.}\ }\textbf
  {\bibinfo {volume} {83}},\ \bibinfo {pages} {3398} (\bibinfo {year}
  {1999})}\BibitemShut {NoStop}%
\bibitem [{\citenamefont {Rodrigues}\ \emph {et~al.}(2019)\citenamefont
  {Rodrigues}, \citenamefont {Bothner},\ and\ \citenamefont
  {Steele}}]{Rodrigues2019}%
  \BibitemOpen
  \bibfield  {author} {\bibinfo {author} {\bibfnamefont {I.~C.}\ \bibnamefont
  {Rodrigues}}, \bibinfo {author} {\bibfnamefont {D.}~\bibnamefont {Bothner}},\
  and\ \bibinfo {author} {\bibfnamefont {G.~A.}\ \bibnamefont {Steele}},\
  }\bibfield  {title} {\bibinfo {title} {Coupling microwave photons to a
  mechanical resonator using quantum interference},\ }\href
  {https://doi.org/10.1038/s41467-019-12964-2} {\bibfield  {journal} {\bibinfo
  {journal} {Nat. Commun.}\ }\textbf {\bibinfo {volume} {10}},\ \bibinfo
  {pages} {1} (\bibinfo {year} {2019})}\BibitemShut {NoStop}%
\bibitem [{\citenamefont {Zoepfl}\ \emph {et~al.}(2020)\citenamefont {Zoepfl},
  \citenamefont {Juan}, \citenamefont {Schneider},\ and\ \citenamefont
  {Kirchmair}}]{Zoepfl_2020}%
  \BibitemOpen
  \bibfield  {author} {\bibinfo {author} {\bibfnamefont {D.}~\bibnamefont
  {Zoepfl}}, \bibinfo {author} {\bibfnamefont {M.~L.}\ \bibnamefont {Juan}},
  \bibinfo {author} {\bibfnamefont {C.~M.~F.}\ \bibnamefont {Schneider}},\ and\
  \bibinfo {author} {\bibfnamefont {G.}~\bibnamefont {Kirchmair}},\ }\bibfield
  {title} {\bibinfo {title} {Single-photon cooling in microwave
  magnetomechanics},\ }\href {https://doi.org/10.1103/PHYSREVLETT.125.023601}
  {\bibfield  {journal} {\bibinfo  {journal} {Phys. Rev. Lett.}\ }\textbf
  {\bibinfo {volume} {125}},\ \bibinfo {pages} {023601} (\bibinfo {year}
  {2020})}\BibitemShut {NoStop}%
\bibitem [{\citenamefont {Schmidt}\ \emph {et~al.}(2020)\citenamefont
  {Schmidt}, \citenamefont {Amawi}, \citenamefont {Pogorzalek}, \citenamefont
  {Deppe}, \citenamefont {Marx}, \citenamefont {Gross},\ and\ \citenamefont
  {Huebl}}]{Schmidt2020}%
  \BibitemOpen
  \bibfield  {author} {\bibinfo {author} {\bibfnamefont {P.}~\bibnamefont
  {Schmidt}}, \bibinfo {author} {\bibfnamefont {M.~T.}\ \bibnamefont {Amawi}},
  \bibinfo {author} {\bibfnamefont {S.}~\bibnamefont {Pogorzalek}}, \bibinfo
  {author} {\bibfnamefont {F.}~\bibnamefont {Deppe}}, \bibinfo {author}
  {\bibfnamefont {A.}~\bibnamefont {Marx}}, \bibinfo {author} {\bibfnamefont
  {R.}~\bibnamefont {Gross}},\ and\ \bibinfo {author} {\bibfnamefont
  {H.}~\bibnamefont {Huebl}},\ }\bibfield  {title} {\bibinfo {title}
  {Sideband-resolved resonator electromechanics based on a nonlinear josephson
  inductance probed on the single-photon level},\ }\href
  {https://doi.org/10.1038/s42005-020-00501-3} {\bibfield  {journal} {\bibinfo
  {journal} {Commun. Phys}\ }\textbf {\bibinfo {volume} {3}},\ \bibinfo {pages}
  {1} (\bibinfo {year} {2020})}\BibitemShut {NoStop}%
\bibitem [{\citenamefont {Bera}\ \emph {et~al.}(2021)\citenamefont {Bera},
  \citenamefont {Majumder}, \citenamefont {Sahu},\ and\ \citenamefont
  {Singh}}]{Bera2021}%
  \BibitemOpen
  \bibfield  {author} {\bibinfo {author} {\bibfnamefont {T.}~\bibnamefont
  {Bera}}, \bibinfo {author} {\bibfnamefont {S.}~\bibnamefont {Majumder}},
  \bibinfo {author} {\bibfnamefont {S.~K.}\ \bibnamefont {Sahu}},\ and\
  \bibinfo {author} {\bibfnamefont {V.}~\bibnamefont {Singh}},\ }\bibfield
  {title} {\bibinfo {title} {Large flux-mediated coupling in hybrid
  electromechanical system with a transmon qubit},\ }\href
  {https://doi.org/10.1038/s42005-020-00514-y} {\bibfield  {journal} {\bibinfo
  {journal} {Commun. Phys. 2021 4:1}\ }\textbf {\bibinfo {volume} {4}},\
  \bibinfo {pages} {1} (\bibinfo {year} {2021})}\BibitemShut {NoStop}%
\bibitem [{\citenamefont {Luschmann}\ \emph {et~al.}(2022)\citenamefont
  {Luschmann}, \citenamefont {Schmidt}, \citenamefont {Deppe}, \citenamefont
  {Marx}, \citenamefont {Sanchez}, \citenamefont {Gross},\ and\ \citenamefont
  {Huebl}}]{Luschmann2022}%
  \BibitemOpen
  \bibfield  {author} {\bibinfo {author} {\bibfnamefont {T.}~\bibnamefont
  {Luschmann}}, \bibinfo {author} {\bibfnamefont {P.}~\bibnamefont {Schmidt}},
  \bibinfo {author} {\bibfnamefont {F.}~\bibnamefont {Deppe}}, \bibinfo
  {author} {\bibfnamefont {A.}~\bibnamefont {Marx}}, \bibinfo {author}
  {\bibfnamefont {A.}~\bibnamefont {Sanchez}}, \bibinfo {author} {\bibfnamefont
  {R.}~\bibnamefont {Gross}},\ and\ \bibinfo {author} {\bibfnamefont
  {H.}~\bibnamefont {Huebl}},\ }\bibfield  {title} {\bibinfo {title}
  {Mechanical frequency control in inductively coupled electromechanical
  systems},\ }\href {https://doi.org/10.1038/s41598-022-05438-x} {\bibfield
  {journal} {\bibinfo  {journal} {Sci. Rep.}\ }\textbf {\bibinfo {volume}
  {12}},\ \bibinfo {pages} {1} (\bibinfo {year} {2022})}\BibitemShut {NoStop}%
\bibitem [{\citenamefont {Zoepfl}\ \emph {et~al.}(2023)\citenamefont {Zoepfl},
  \citenamefont {Juan}, \citenamefont {{Diaz-Naufal}}, \citenamefont
  {Schneider}, \citenamefont {Deeg}, \citenamefont {Sharafiev}, \citenamefont
  {Metelmann},\ and\ \citenamefont
  {Kirchmair}}]{zoepflKerrEnhancedBackaction2023}%
  \BibitemOpen
  \bibfield  {author} {\bibinfo {author} {\bibfnamefont {D.}~\bibnamefont
  {Zoepfl}}, \bibinfo {author} {\bibfnamefont {M.~L.}\ \bibnamefont {Juan}},
  \bibinfo {author} {\bibfnamefont {N.}~\bibnamefont {{Diaz-Naufal}}}, \bibinfo
  {author} {\bibfnamefont {C.~M.~F.}\ \bibnamefont {Schneider}}, \bibinfo
  {author} {\bibfnamefont {L.~F.}\ \bibnamefont {Deeg}}, \bibinfo {author}
  {\bibfnamefont {A.}~\bibnamefont {Sharafiev}}, \bibinfo {author}
  {\bibfnamefont {A.}~\bibnamefont {Metelmann}},\ and\ \bibinfo {author}
  {\bibfnamefont {G.}~\bibnamefont {Kirchmair}},\ }\bibfield  {title} {\bibinfo
  {title} {Kerr {{Enhanced Backaction Cooling}} in {{Magnetomechanics}}},\
  }\href {https://doi.org/10.1103/PhysRevLett.130.033601} {\bibfield  {journal}
  {\bibinfo  {journal} {Phys. Rev. Lett.}\ }\textbf {\bibinfo {volume} {130}},\
  \bibinfo {pages} {033601} (\bibinfo {year} {2023})}\BibitemShut {NoStop}%
\bibitem [{\citenamefont {Gieseler}\ \emph {et~al.}(2013)\citenamefont
  {Gieseler}, \citenamefont {Novotny},\ and\ \citenamefont
  {Quidant}}]{gieselerThermalNonlinearitiesNanomechanical2013}%
  \BibitemOpen
  \bibfield  {author} {\bibinfo {author} {\bibfnamefont {J.}~\bibnamefont
  {Gieseler}}, \bibinfo {author} {\bibfnamefont {L.}~\bibnamefont {Novotny}},\
  and\ \bibinfo {author} {\bibfnamefont {R.}~\bibnamefont {Quidant}},\
  }\bibfield  {title} {\bibinfo {title} {Thermal nonlinearities in a
  nanomechanical oscillator},\ }\href {https://doi.org/10.1038/nphys2798}
  {\bibfield  {journal} {\bibinfo  {journal} {Nat. Phys.}\ }\textbf {\bibinfo
  {volume} {9}},\ \bibinfo {pages} {806} (\bibinfo {year} {2013})}\BibitemShut
  {NoStop}%
\bibitem [{\citenamefont {Gieseler}\ \emph {et~al.}(2014)\citenamefont
  {Gieseler}, \citenamefont {Spasenovic}, \citenamefont {Novotny},\ and\
  \citenamefont {Quidant}}]{Gieseler2014}%
  \BibitemOpen
  \bibfield  {author} {\bibinfo {author} {\bibfnamefont {J.}~\bibnamefont
  {Gieseler}}, \bibinfo {author} {\bibfnamefont {M.}~\bibnamefont
  {Spasenovic}}, \bibinfo {author} {\bibfnamefont {L.}~\bibnamefont
  {Novotny}},\ and\ \bibinfo {author} {\bibfnamefont {R.}~\bibnamefont
  {Quidant}},\ }\bibfield  {title} {\bibinfo {title} {Nonlinear mode coupling
  and synchronization of a vacuum-trapped nanoparticle},\ }\href
  {https://doi.org/10.1103/PHYSREVLETT.112.103603/FIGURES/3/MEDIUM} {\bibfield
  {journal} {\bibinfo  {journal} {Phys. Rev. Lett.}\ }\textbf {\bibinfo
  {volume} {112}},\ \bibinfo {pages} {103603} (\bibinfo {year}
  {2014})}\BibitemShut {NoStop}%
\bibitem [{\citenamefont {Setter}\ \emph {et~al.}(2019)\citenamefont {Setter},
  \citenamefont {Vovrosh},\ and\ \citenamefont {Ulbricht}}]{Setter2019}%
  \BibitemOpen
  \bibfield  {author} {\bibinfo {author} {\bibfnamefont {A.}~\bibnamefont
  {Setter}}, \bibinfo {author} {\bibfnamefont {J.}~\bibnamefont {Vovrosh}},\
  and\ \bibinfo {author} {\bibfnamefont {H.}~\bibnamefont {Ulbricht}},\
  }\bibfield  {title} {\bibinfo {title} {Characterization of non-linearities
  through mechanical squeezing in levitated optomechanics},\ }\href
  {https://doi.org/10.1063/1.5116121} {\bibfield  {journal} {\bibinfo
  {journal} {Applied Physics Letters}\ }\textbf {\bibinfo {volume} {115}},\
  \bibinfo {pages} {153106} (\bibinfo {year} {2019})}\BibitemShut {NoStop}%
\bibitem [{\citenamefont {Zheng}\ \emph
  {et~al.}(2020{\natexlab{b}})\citenamefont {Zheng}, \citenamefont {Zhou},
  \citenamefont {Dong}, \citenamefont {Qiu}, \citenamefont {Chen},
  \citenamefont {Guo},\ and\ \citenamefont {Sun}}]{Zheng2020}%
  \BibitemOpen
  \bibfield  {author} {\bibinfo {author} {\bibfnamefont {Y.}~\bibnamefont
  {Zheng}}, \bibinfo {author} {\bibfnamefont {L.-M.}\ \bibnamefont {Zhou}},
  \bibinfo {author} {\bibfnamefont {Y.}~\bibnamefont {Dong}}, \bibinfo {author}
  {\bibfnamefont {C.-W.}\ \bibnamefont {Qiu}}, \bibinfo {author} {\bibfnamefont
  {X.-D.}\ \bibnamefont {Chen}}, \bibinfo {author} {\bibfnamefont {G.-C.}\
  \bibnamefont {Guo}},\ and\ \bibinfo {author} {\bibfnamefont {F.-W.}\
  \bibnamefont {Sun}},\ }\bibfield  {title} {\bibinfo {title} {Robust
  optical-levitation-based metrology of nanoparticle's position and mass},\
  }\href {https://doi.org/10.1103/PhysRevLett.124.223603} {\bibfield  {journal}
  {\bibinfo  {journal} {Phys. Rev. Lett.}\ }\textbf {\bibinfo {volume} {124}},\
  \bibinfo {pages} {223603} (\bibinfo {year} {2020}{\natexlab{b}})}\BibitemShut
  {NoStop}%
\bibitem [{\citenamefont {Flaj{\v s}manov{\'a}}\ \emph
  {et~al.}(2020)\citenamefont {Flaj{\v s}manov{\'a}}, \citenamefont {{\v
  S}iler}, \citenamefont {Jedli{\v c}ka}, \citenamefont {Hrub{\'y}},
  \citenamefont {Brzobohat{\'y}}, \citenamefont {Filip},\ and\ \citenamefont
  {Zem{\'a}nek}}]{Flajsmanova2020}%
  \BibitemOpen
  \bibfield  {author} {\bibinfo {author} {\bibfnamefont {J.}~\bibnamefont
  {Flaj{\v s}manov{\'a}}}, \bibinfo {author} {\bibfnamefont {M.}~\bibnamefont
  {{\v S}iler}}, \bibinfo {author} {\bibfnamefont {P.}~\bibnamefont {Jedli{\v
  c}ka}}, \bibinfo {author} {\bibfnamefont {F.}~\bibnamefont {Hrub{\'y}}},
  \bibinfo {author} {\bibfnamefont {O.}~\bibnamefont {Brzobohat{\'y}}},
  \bibinfo {author} {\bibfnamefont {R.}~\bibnamefont {Filip}},\ and\ \bibinfo
  {author} {\bibfnamefont {P.}~\bibnamefont {Zem{\'a}nek}},\ }\bibfield
  {title} {\bibinfo {title} {Using the transient trajectories of an optically
  levitated nanoparticle to characterize a stochastic duffing oscillator},\
  }\href {https://doi.org/10.1038/s41598-020-70908-z} {\bibfield  {journal}
  {\bibinfo  {journal} {Scientific Reports 2020 10:1}\ }\textbf {\bibinfo
  {volume} {10}},\ \bibinfo {pages} {1} (\bibinfo {year} {2020})}\BibitemShut
  {NoStop}%
\bibitem [{\citenamefont {Maisonobe}\ \emph {et~al.}(2018)\citenamefont
  {Maisonobe}, \citenamefont {Billard}, \citenamefont {Jesus}, \citenamefont
  {Juillard}, \citenamefont {Misiak}, \citenamefont {Olivieri}, \citenamefont
  {Sayah},\ and\ \citenamefont {Vagneron}}]{Maisonobe2018}%
  \BibitemOpen
  \bibfield  {author} {\bibinfo {author} {\bibfnamefont {R.}~\bibnamefont
  {Maisonobe}}, \bibinfo {author} {\bibfnamefont {J.}~\bibnamefont {Billard}},
  \bibinfo {author} {\bibfnamefont {M.~D.}\ \bibnamefont {Jesus}}, \bibinfo
  {author} {\bibfnamefont {A.}~\bibnamefont {Juillard}}, \bibinfo {author}
  {\bibfnamefont {D.}~\bibnamefont {Misiak}}, \bibinfo {author} {\bibfnamefont
  {E.}~\bibnamefont {Olivieri}}, \bibinfo {author} {\bibfnamefont
  {S.}~\bibnamefont {Sayah}},\ and\ \bibinfo {author} {\bibfnamefont
  {L.}~\bibnamefont {Vagneron}},\ }\bibfield  {title} {\bibinfo {title}
  {Vibration decoupling system for massive bolometers in dry cryostats},\
  }\href {https://doi.org/10.1088/1748-0221/13/08/t08009} {\bibfield  {journal}
  {\bibinfo  {journal} {J. Instrum.}\ }\textbf {\bibinfo {volume} {13}}\bibinfo
   {number} { (08)},\ \bibinfo {pages} {T08009}}\BibitemShut {NoStop}%
\bibitem [{\citenamefont {Wit}\ \emph {et~al.}(2019)\citenamefont {Wit},
  \citenamefont {Welker}, \citenamefont {Heeck}, \citenamefont {Buters},
  \citenamefont {Eerkens}, \citenamefont {Koning}, \citenamefont {der Meer},
  \citenamefont {Bouwmeester},\ and\ \citenamefont {Oosterkamp}}]{DeWit2019}%
  \BibitemOpen
\bibfield  {number} {  }\bibfield  {author} {\bibinfo {author} {\bibfnamefont
  {M.~D.}\ \bibnamefont {Wit}}, \bibinfo {author} {\bibfnamefont
  {G.}~\bibnamefont {Welker}}, \bibinfo {author} {\bibfnamefont
  {K.}~\bibnamefont {Heeck}}, \bibinfo {author} {\bibfnamefont {F.~M.}\
  \bibnamefont {Buters}}, \bibinfo {author} {\bibfnamefont {H.~J.}\
  \bibnamefont {Eerkens}}, \bibinfo {author} {\bibfnamefont {G.}~\bibnamefont
  {Koning}}, \bibinfo {author} {\bibfnamefont {H.~V.}\ \bibnamefont {der
  Meer}}, \bibinfo {author} {\bibfnamefont {D.}~\bibnamefont {Bouwmeester}},\
  and\ \bibinfo {author} {\bibfnamefont {T.~H.}\ \bibnamefont {Oosterkamp}},\
  }\bibfield  {title} {\bibinfo {title} {Vibration isolation with high thermal
  conductance for a cryogen-free dilution refrigerator},\ }\href
  {https://doi.org/10.1063/1.5066618} {\bibfield  {journal} {\bibinfo
  {journal} {Rev. Sci. Instrum.}\ }\textbf {\bibinfo {volume} {90}},\ \bibinfo
  {pages} {015112} (\bibinfo {year} {2019})}\BibitemShut {NoStop}%
\bibitem [{\citenamefont {Rossi}\ \emph {et~al.}(2018)\citenamefont {Rossi},
  \citenamefont {Mason}, \citenamefont {Chen}, \citenamefont {Tsaturyan},\ and\
  \citenamefont {Schliesser}}]{rossiMeasurementbasedQuantumControl2018}%
  \BibitemOpen
  \bibfield  {author} {\bibinfo {author} {\bibfnamefont {M.}~\bibnamefont
  {Rossi}}, \bibinfo {author} {\bibfnamefont {D.}~\bibnamefont {Mason}},
  \bibinfo {author} {\bibfnamefont {J.}~\bibnamefont {Chen}}, \bibinfo {author}
  {\bibfnamefont {Y.}~\bibnamefont {Tsaturyan}},\ and\ \bibinfo {author}
  {\bibfnamefont {A.}~\bibnamefont {Schliesser}},\ }\bibfield  {title}
  {\bibinfo {title} {Measurement-based quantum control of mechanical motion},\
  }\href {https://doi.org/10.1038/s41586-018-0643-8} {\bibfield  {journal}
  {\bibinfo  {journal} {Nature}\ }\textbf {\bibinfo {volume} {563}},\ \bibinfo
  {pages} {53} (\bibinfo {year} {2018})}\BibitemShut {NoStop}%
\bibitem [{\citenamefont {Hofer}\ and\ \citenamefont
  {Aspelmeyer}(2019)}]{Hofer_2019}%
  \BibitemOpen
  \bibfield  {author} {\bibinfo {author} {\bibfnamefont {J.}~\bibnamefont
  {Hofer}}\ and\ \bibinfo {author} {\bibfnamefont {M.}~\bibnamefont
  {Aspelmeyer}},\ }\bibfield  {title} {\bibinfo {title} {Analytic solutions to
  the maxwell{\textendash}london equations and levitation force for a
  superconducting sphere in a quadrupole field},\ }\href
  {https://doi.org/10.1088/1402-4896/ab0c44} {\bibfield  {journal} {\bibinfo
  {journal} {Phys. Scr.}\ }\textbf {\bibinfo {volume} {94}},\ \bibinfo {pages}
  {125508} (\bibinfo {year} {2019})}\BibitemShut {NoStop}%
\bibitem [{\citenamefont {Bachtold}\ \emph {et~al.}(2022)\citenamefont
  {Bachtold}, \citenamefont {Moser},\ and\ \citenamefont
  {Dykman}}]{bachtoldMesoscopicPhysicsNanomechanical2022}%
  \BibitemOpen
  \bibfield  {author} {\bibinfo {author} {\bibfnamefont {A.}~\bibnamefont
  {Bachtold}}, \bibinfo {author} {\bibfnamefont {J.}~\bibnamefont {Moser}},\
  and\ \bibinfo {author} {\bibfnamefont {M.~I.}\ \bibnamefont {Dykman}},\
  }\bibfield  {title} {\bibinfo {title} {Mesoscopic physics of nanomechanical
  systems},\ }\href {https://doi.org/10.1103/RevModPhys.94.045005} {\bibfield
  {journal} {\bibinfo  {journal} {Rev. Mod. Phys.}\ }\textbf {\bibinfo {volume}
  {94}},\ \bibinfo {pages} {045005} (\bibinfo {year} {2022})}\BibitemShut
  {NoStop}%
\bibitem [{\citenamefont {Drung}\ \emph {et~al.}(2007)\citenamefont {Drung},
  \citenamefont {Abmann}, \citenamefont {Beyer}, \citenamefont {Kirste},
  \citenamefont {Peters}, \citenamefont {Ruede},\ and\ \citenamefont
  {Schurig}}]{Drung2007}%
  \BibitemOpen
  \bibfield  {author} {\bibinfo {author} {\bibfnamefont {D.}~\bibnamefont
  {Drung}}, \bibinfo {author} {\bibfnamefont {C.}~\bibnamefont {Abmann}},
  \bibinfo {author} {\bibfnamefont {J.}~\bibnamefont {Beyer}}, \bibinfo
  {author} {\bibfnamefont {A.}~\bibnamefont {Kirste}}, \bibinfo {author}
  {\bibfnamefont {M.}~\bibnamefont {Peters}}, \bibinfo {author} {\bibfnamefont
  {F.}~\bibnamefont {Ruede}},\ and\ \bibinfo {author} {\bibfnamefont
  {T.}~\bibnamefont {Schurig}},\ }\bibfield  {title} {\bibinfo {title} {Highly
  sensitive and easy-to-use squid sensors},\ }\href
  {https://doi.org/10.1109/TASC.2007.897403} {\bibfield  {journal} {\bibinfo
  {journal} {IEEE Trans. Appl. Supercond.}\ }\textbf {\bibinfo {volume} {17}},\
  \bibinfo {pages} {699} (\bibinfo {year} {2007})}\BibitemShut {NoStop}%
\end{thebibliography}%

\end{document}